\documentclass[preprint]{elsarticle}

\usepackage{amssymb}
\usepackage{graphicx}
\usepackage{caption}
\usepackage{pdfpages}
\usepackage{adjustbox}
\usepackage{rotating}
\usepackage{floatpag}
\usepackage{wrapfig}
\usepackage[T1]{fontenc}
\usepackage{mathrsfs }
\usepackage{multirow}
\usepackage{hhline}
\usepackage{enumerate}
\usepackage{enumitem}
\usepackage{adjustbox}
\usepackage[lined,commentsnumbered]{algorithm2e}
\usepackage{booktabs}
\usepackage{mathtools}

\usepackage{marvosym}

\biboptions{sort&compress}

\usepackage[english]{babel}
\usepackage[font=small,labelfont=bf]{caption}
\usepackage[tight,footnotesize]{subfigure}

\newlength\marincrease


\usepackage{makecell}

\usepackage{etoolbox}

\usepackage[bookmarks=true,
bookmarksnumbered=false, 
bookmarksopen=false, 
colorlinks=true,
linkcolor=webred]{hyperref}
\definecolor{webgreen}{rgb}{0, 0.5, 0} 
\definecolor{webblue}{rgb}{0, 0, 0.5} 
\definecolor{webred}{rgb}{0.5, 0, 0} 
\definecolor{webblack}{rgb}{0, 0, 0} 

\makeatletter

\let\old@@children\@@children
\def\@@children{\futurelet\my@next\my@@children}
\def\my@@children{%
\ifx\my@next\missing\else
\expandafter\@gobble
\fi
\expandafter\old@@children}

\makeatother

\newcommand{\missing}{ \edge[draw=none]; {} }

\journal{Elsevier}

\begin{document}

\renewcommand{\thesection}{\Roman{section}}

\renewcommand{\thesection}{\Roman{section}}
\newtheorem{theorem}{Theorem}
\newtheorem{proposition}{Proposition}
\newtheorem{eqed}{Example}
\newtheorem {lemmaa}{Lemma}
\newtheorem {observation}{Observation}
\newtheorem {example}{Example}
\newtheorem {corollary}{Corollary}
\newtheorem {defnn}{Definition}
\newenvironment{proof}{\noindent {\bf Proof :\ } }{\hfill$\Box$ }
\newtheorem {con}{Condition}
\newtheorem {conjecturee}{Conjecture}
\newtheorem {procd}{Procedure}
\newtheorem {rules}{Rule}

\newenvironment{proof of correctness}{\noindent {\bf Reason :\ } }{\hfill$\Box$ }
\newenvironment{definition}[1][Definition]{\begin{defnn} \sl}{\end{defnn}}
\newenvironment{algo}{\begin{algorithm}}{\end{algorithm}}

\begin{frontmatter}

\title{Cellular Automata Model for Non-Structural Proteins Comparing Transmissibility and Pathogenesis of SARS Covid (CoV-2, CoV) and MERS Covid}

\author[mymainaddress1]{Raju Hazari\corref{mycorrespondingauthor}}
\cortext[mycorrespondingauthor]{Corresponding author}
\ead{hazariraju0201@gmail.com }

\address[mymainaddress1]{Department of Computer Science and Engineering, National Institute of Technology Calicut, Kerala, India 673601}

\author[mymainaddress2]{Parimal Pal Chaudhuri}
\ead{ppc@carlbio.com}

\address[mymainaddress2]{Retired Professor, Indian Institute of Technology Kharagpur, India 721302}

\begin{abstract}

Significantly higher transmissibility of SARS CoV-2 (2019) compared to SARS CoV (2003) can be attributed to mutations of structural proteins (Spike S, Nucleocapsid N, Membrane M, and Envelope E) and the role played by non-structural proteins (nsps) and accessory proteins (ORFs) for viral replication, assembly and shedding. The non-structural proteins (nsps) avail host protein synthesis machinery to initiate viral replication, along with neutralization of host immune defense. The key protein out of the 16 nsps, is the non-structural protein nsp1, also known as the leader protein. Nsp1 leads the process of hijacking host resources by blocking host translation. This paper concentrates on the analysis of nsps of SARS covid (CoV-2, CoV) and MERS covid based on Cellular Automata enhanced Machine Learning (CAML) model developed for study of biological strings. This computational model compares deviation of structure - function of CoV-2 from that of CoV employing CAML model parameters derived out of CA evolution of amino acid chains of nsps. This comparative analysis points to - (i) higher transmissibility of CoV-2 compared to CoV for major nsps, and (ii) deviation of MERS covid from SARS CoV in respect of virulence and pathogenesis. A Machine Learning (ML) framework has been designed to map the CAML model parameters to the physical domain features reported in in-vitro/in-vivo/in-silico experimental studies. The ML framework enables us to learn the permissible range of model parameters derived out of mutational study of sixteen nsps of three viruses.

\end{abstract}


\end{frontmatter}

\section{Introduction}
\label{intro}

Twin objectives of a living organism are - to survive and replicate. Virus, a cellular parasite, fulfils these objectives by infecting cells of a living organism for its replication and transmission. In order to achieve this goal, a virus typically encodes proteins similar to those of host proteins and/or develop strategy to avail the services of host cell. Intimate relationship of virus with host and different aspects of viral symbiosis are reported in \cite{roossinck2017symbiosis}. A recent survey on human microbiome and CoV-2 relationship is reported in \cite{yamamoto2021human}.                                    In the context of on going pandemic due to SARS CoV-2, it is absolutely essential to decipher structure and function of its four structural proteins (Spike S, Nucleocapsid N, Membrane M, and Envelope E), 16 non-structural proteins (nsps), and 9 accessory proteins (ORFs). Different variants of CoV-2 evolving around the globe with higher transmissibility and/or virulence is a major concern for public health and local/global economy. Consequently, a large number of recent publications have concentrated on study of different aspects of structural proteins, nsps and ORFs. Design of vaccine, drug, anti-viral agents are also reported in many publications. The immune system of a living organism is a complex network evolved to protect a species from invading pathogens. Complexity of human immune system is significantly higher. On sensing an invading pathogen, immune system adjusts its pathways to fight a virus. However, a virus has evolved strategy for its survival and replication in the host by evading immune system defense. Different aspects of host immune response and viral immune evasion are covered by many authors.

The focus of the current paper is non-structural proteins (nsps) for investigating their contribution for significantly higher transmissibility of CoV-2 compared to CoV, and higher pathogenesis of MERS covid. A brief survey on nsps follows in Section~\ref{survey_nsps}. Prior to introduction of CAML (Cellular Automata enhanced Machine Learning) model, Cellular Automata (CA) preliminaries are presented in Section~\ref{CA_preli}. The CAML model has been designed in two phases:                                                                                                                                                                              Phase I - Bottom-up CA rule design based on atomic structure of amino acids; and                                                    Phase II - Top-down Machine Learning (ML) framework design.\\ Cellular Automata enhanced Machine Learning (CAML) is a new kind of Machine Learning model in which ML enhances characterization of CA evolution patterns. While CA rule design of Phase I is reported in Section~\ref{Sec:CA_rule_for_AA}, the background and design of the ML framework are covered in Section~\ref{CA_Evolution_CL_Graph} and \ref{Generic_CAML}. The experimental results derived out of CAML model for major nsps of CoV-2, CoV, and MERS are presented in Section~\ref{result_nsp1} and \ref{result_other_nsps}. The results reported clearly establish contribution of nsps towards higher transmissibility of CoV-2 compared to CoV. A brief survey on nsps follows.

\section{SARS Covid and MERS Covid nsps - A Brief Survey}
\label{survey_nsps}

Based on the survey of published literature, we have divided sixteen nsps into two groups. Group1 covers ten nsps which play primary role for viral replication - nsp1, nsp3, nsp5, nsp6, nsp8, nsp10, nsp12, nsp14, nsp15, nsp16. Group2 covers remaining 6 nsps (nsp2, nsp4, nsp7, nsp9, nsp11, nsp13) which play secondary role in corona virus life cycle. Group1 nsps can be further subdivided in to enzymatic and non-enzymatic proteins. All the eight proteins except nsp1 and nsp6 are enzymes having specific catalytic sites and other functionally important AAs associated with enzymatic function. By contrast, nsp1 binds on other biomolecules - host ribosome and 5'UTR of viral RNA, while nsp6 interacts with host proteins of autophagy signaling pathways for viral replication.

A brief survey on the tasks associated with function of each Group1 nsp is next presented. The survey output serves as the input for `Design of Machine Learning (ML) Platform' reported in Section~\ref{ML_Framework_Design}. In the rest of this paper, an amino acid is referred to as AA and the acronyms NTD and CTD refer to N-Terminal Domain and C-Terminal Domain of a protein chain respectively.

\subsection{Tasks executed by nsp1}
\label{Task_nsp1}

\vspace{2mm}
In view of its leading position on viral RNA, nsp1 is often referred to as the leader protein. Subsequent to entry in host cell, viral pathogenesis proceeds by executing the following tasks reported in-vivo/in-vitro experimental studies \cite{shen2021lysine, schubert2020sars, nakagawa2021mechanisms, simeoni2021nsp1, terada2017mers, benedetti2020emerging, clark2021structure, kumar2020sars, tidu2021viral, vankadari2020structure, de2020translational, miao2021secondary, tanaka2012severe, yang2015structure, mohammadi2021transcription, kikkert2020innate, thoms2020structural, taefehshokr2020covid, lokugamage2015middle}.  Both CoV-2 and CoV nsp1 have 180 AA, while MERS nsp1 has 193 AA. MERS covid differs significantly from SARS covid in respect of its interaction with other biomolecules. Detailed analysis of MERS nsp1 is reported in \cite{terada2017mers}. The results derived out of CAML model are presented in Section~\ref{result_nsp1} (SARS covid in Section~\ref{nsp1_Task_Ta1}, \ref{nsp1_Task_Ta2}, \ref{nsp1_Task_Ta3} and MERS covid in Section~\ref{MERS_nsp1_Result}).  In subsequent discussions, the task X (X = 1, 2, 3) executed by nsp1 is referred to as TaX.

\vspace{2mm}
\noindent
\textbf{Task Ta1:} The very first task of nsp1 of SARS covid (CoV, CoV2) is to hijack host translation machinery to shut off host mRNA translation. Implementation of Ta1 proceeds through binding of nsp1 to ribosomal subunit S40. Details of Ta1 execution has been dealt with by a number of authors \cite{shen2021lysine, schubert2020sars, nakagawa2021mechanisms, simeoni2021nsp1, terada2017mers, benedetti2020emerging, clark2021structure, kumar2020sars, tidu2021viral, vankadari2020structure}. Two AAs of KH motif in CTD (K164, H165) of CoV-2 and CoV nsp1 interact with host ribosome to inhibit binding of host mRNA on ribosome necessary for protein synthesis. The amino acid K164 has been reported to be the critical AA of nsp1 for binding with host ribosome \cite{shen2021lysine}. Further, it has been reported that the AA (154 to 160) and (166 to 179) containing two helices are inserted in the entry point of ribosomal RNA. Section~\ref{nsp1_Task_Ta1} reports the results derived out of CAML model for task Ta1 on comparing CoV-2 and CoV nsp1 covering the critical KH motif (K164, H165).                                                                                                                                                                                               On aligning MERS sequence with that of CoV-2/CoV, a number of authors have pointed to the AA location K181 for host translation shut off through binding of MERS nsp1 on ribosome.  Section~\ref{MERS_nsp1_Result} reports results derived out of CAML  model for execution of task Ta1 of MERS nsp1.

\vspace{2mm}
\noindent
\textbf{Task Ta2:} The next task (referred to as Ta2) of nsp1 is to ensure that nsp1 mediated host translation shut off does not affect viral RNA translation. Uninterrupted viral RNA translation through viral gene expression is ensured through interaction of nsp1 with Stem Loop 1 (SL1) of 5'UTR secondary structure of viral RNA. The overall secondary structure of 5'UTR for SARS CoV-2 and COV is conserved. Execution of the task Ta2 is covered in many publications \cite{vankadari2020structure, de2020translational, miao2021secondary, tanaka2012severe, yang2015structure, mohammadi2021transcription}. Interaction of the 5'UTR with nsp1 through R124 and K125 is reported in \cite{tanaka2012severe} - such interaction is required for evasion of nsp1-mediated translational suppression. MERS covid nsp1 (with 193 amino acids) differs from SARS CoV (with 180 amino acids) in respect of viral RNA recognition. Comparison of different features of MERS covid and SARS covid is covered in \cite{terada2017mers} both in respect of Ta1 and Ta2.                                                                                                                                                                            Section~\ref{nsp1_Task_Ta2} reports results derived out of CAML model on CoV-2 and CoV nsp1 on task Ta2 considering the critical AA of RK motif (R124, K125). Through successive deletion experiment, the interaction region of MERS nsp1 with 5'UTR has been identified as AA location 11 to 15 \cite{terada2017mers}, while the authors in \cite{terada2017mers, de2020translational} have also highlighted the importance amino acid R13 of MERS nsp1 for recognition of RNA 5'UTR. Section~\ref{MERS_nsp1_Result} reports the results derived out of CAML model for execution of task Ta2.

\vspace{2mm}
\noindent
\textbf{Task Ta3:} The third task (Ta3) of nsp1 is to inhibit/restrict production of host defence immune system components - Interferons (INF-I, INF-gamma, and Interleukin). The task Ta3 can be assumed to be a corollary to Ta1, since blockage of host mRNA translation stops synthesis of host proteins necessary to activate immune pathways \cite{lokugamage2015middle}. Virus also blocks export of mRNA from nucleus of the host cell to cytoplasm where some of the host proteins are made. A number of authors have concentrated on how SARS and MARS covid evade host immune response; a representative set is noted in \cite{mohammadi2021transcription, kikkert2020innate, thoms2020structural, taefehshokr2020covid, lokugamage2015middle}.
It has been reported \cite{terada2017mers, lokugamage2015middle} that MERS nsp1 displays host mRNA cleavage function. Two functions of MERS nsp1 - (i) host translational shut off, and (ii) host mRNA cleavage are executed by two different regions. While host translational shut off is executed by the CTD AA K179/K181, the task of mRNA cleavage is executed by the AA pair (R146, K147). As a result, the proteins necessary for activation of host immune response are not synthesized leading to compromised host immune response due to MERS nsp1.
Section~\ref{nsp1_Task_Ta3} reports the results derived out of CAML model for execution of task Ta3 of CoV-2 and CoV nsp1, while the results for MERS nsp1 are reported in Section~\ref{MERS_nsp1_Result}.

\vspace{2mm}
\noindent
\emph{Observation1 from Analysis of nsp sequences and Published Experimental Results :} From the analysis of SARS and MERS covid nsps and survey of published in vitro/in vivo experimental results, it can be observed that polar positive AAs (Lys K, Arg R, His H) of nsp1, play a prominent role for interaction with host proteins and viral 5'UTR. CAML model results reported in Section~\ref{result_nsp1} are derived on employing this key observation on nsp1 AA sequence.

\subsection{Tasks executed by nsp3}
\label{Task_nsp3}

\vspace{2mm}
The nsp3 is the largest element of the RTC (Replication/Transcription Complex). A number of authors \cite{lei2018nsp3, shin2020papain, stasiulewicz2021sars, osipiuk2021structure, armstrong2021biochemical, gao2021crystal, bosken2020insights, baez2015sars, shen2021potent} have reported different functional aspects of nsp3. In addition to cleaving, nsp3 interacts with host proteins to downgrade host innate immune response. The papain-like protease domain (PLpro), encoded within nsp3 (Figure 1(a) in \cite{armstrong2021biochemical}), executes the cleavage of individual nsp (nsp1, nsp2, nsp3, nsp4) from viral RNA. Nsp3 also contains other domain at its CTD and NDT for execution of multiple functions. CoV-2 PLpro has been reported to attenuate type I interferon responses \cite{shin2020papain}; inhibition of PLpro reduces viral replication due to normal function of antiviral interferon pathway. Location of catalytic triads for PLpro are confirmed as (C111 - H272 - D286); other catalytically important amino acids are  (W93, W106, D108, N109 for CoV-2, CoV, and MERS),  with W106  replaced by L106 in MERS. In view of its important role for viral replication, PLpro has been identified as an attractive therapeutic target for CoV-2.  A number of authors \cite{stasiulewicz2021sars, osipiuk2021structure, armstrong2021biochemical, bosken2020insights, shen2021potent} have concentrated on design of new or refurbished drugs (out of CoV PLpro inhibitor) to bind on CoV-2 PLpro to make it inactive.

MD simulation results presented in \cite{bosken2020insights} characterizes dynamics of selected drug binding to identify similarities and differences among PLpro of CoV-2, CoV, and MERS. As per their report, CoV2 PLpro overall dynamics are similar to those of CoV PLpro with identical hydrophobic cleft region around P247, P248, Y264. Based on the  brief survey, two tasks executed by PLpro can be summarized as -\\                                                                                                                                                 (i) cleavage of viral RNA to generate nsps (nsp1, nsp2, nsp3, nsp4); and\\                                                                                                                                                       (ii) downgrade host immune response.

\noindent 
Functionally important and conserved amino acids in PLpro reported  in [22 - 30] across three viruses  CoV-2, CoV, and MERS are as follows -\\                                                                                                                                                                                                                                                   (a) Catalytic triads (C111, H272, and D286);\\                                                                                                                                                                                                   (b) other catalytically important residues : W93, W106, D108, N109 (with L106 in MERS; and\\ 
(c) hydrophobic cleft (P247, P248, Y264) common for CoV-2 and CoV.\\                                                                                                                 Section~\ref{CAML_nsp3_result} reports a comparative study of CoV-2 and CoV nsp3 results derived out of CAML model considering the AAs listed under (a), (b), and (c).

\subsection{Task executed by nsp5}
\label{Task_nsp5}

\vspace{2mm}
Nsp5 (MPro, 3CLPro) is a protease crucial for RNA replication. The name `main protease' (or Mpro) refers to the critical role of this protease for coronavirus gene expression and replicase processing; it is also referred to `3C- like protease' (3CLpro) due to similarities between this protease and 3C proteases of picornaviruses. Different aspects of nsp5 function are covered in \cite{citarella2021sars, grottesi2020computational, ferreira2020biochemical, gossen2021blueprint, lee2020crystallographic, miczi2020identification, jin2020structure, roe2021targeting}.

A protease, in general, speeds up proteolysis - the function of breaking down a protein chain into smaller polypeptides by cleaving the peptide bonds between a pair of AA in the chain. Detailed structural analysis of nsp5 along with sequential execution of nsp5 task are reported in \cite{roe2021targeting}. The nsp5 AA chain covers 3 Domains (Domain DI- 1 to 100, DII- 101 to 199, and DIII- 200 to 306) - the N-terminal covers DI and DII (1 to 199). Around 96\% of AA chain are identical for CoV-2 and CoV with difference in 12 AA - majority of these AA are located away from Active site. The consensus cleavage Site (catalytic dyed residues) are H41 (in Domain 1) and C145 (in Domain 2). The substrate-binding site is located in a pocket between domain I and II. The catalytic pocket is enclosed by two loop regions (AA 44 to 53 and AA 184 to 193) that physically occlude the path to the catalytic site and are known to play an important role in the catalysis. It has been reported \cite{roe2021targeting} that nsp5 is anchored to nsp4 and nsp6 to form a dimer. Strong propensity for dimerization, it seems, is a necessity for execution of sequential cleavage of different viral proteins. The dynamics of monomerized and dimerized conformation differs, as confirmed through MD simulation \cite{grottesi2020computational}. We assume that dimerization and subsequent sequential cleavage of viral proteins for CoV-2 and CoV nsp5 are identical.  Under this assumption we compare the structure-function of CoV-2 and CoV nsp5 monomer employing CAML model for execution of task executed by nsp5: breaking down viral polyprotein by cleaving its functional proteins (structural, non-structural, accessory proteins), that is a prerequisite for virus replication subsequent to its entry in host cell.  

In Section~\ref{CAML_nsp5_result} we report a comparative study of CoV and CoV-2 nsp5 monomer in respect of execution of the overall nsp5 tasks.

\subsection{Task executed by nsp6}
\label{Task_nsp6}

\vspace{2mm}
The term `autophagy' (self-eating) was coined by Belgian Noble Laurate Christian de Duve to describe this process subsequent to discovery of the associated biomolecule lysosome. Autophagosome is the key biomolecule for degradation of damaged/abnormal cytoplasmic contents. Different aspects of nsp6 function are covered in \cite{cottam2014coronavirus, benvenuto2020evolutionary, miller2020coronavirus, choi2018autophagy, santerre2021sars, schmidt2021sars, steele2015role, snijder2020unifying}. The nsp6 plays a key role in the initial induction of autophagosomes from host reticulum endoplasmic for transporting such components in sack-like vesicles transported to the lysosome for their destruction. At a later stage nsp6 limits the expansion of these phagosomes that are no longer able to deliver viral components to lysosomes \cite{cottam2014coronavirus, benvenuto2020evolutionary, choi2018autophagy}. In the process, the virus maintains a delicate balance in autophagy signaling to escape its destruction, while availing the benefits of host autophagy that provides structural and nutrient/energy support \cite{steele2015role}.

NSP6 structure has six transmembrane helices and a highly conserved C-terminus. It is inserted in the ER membrane along with Nsp3 and Nsp4 multi-pass transmembrane proteins during the assembly of virus replication complex to form Double Membrane Vesicles (DMVs). The viral proteins cause stress on several cell organelles leading to the formation of double-membrane vesicles (DMVs). The virus employs DMV to hide and translate its accessory proteins. DMVs or membrane rearrangements can be viewed as extrusion of the ER membrane \cite{cottam2014coronavirus, benvenuto2020evolutionary, miller2020coronavirus, choi2018autophagy, santerre2021sars, schmidt2021sars, steele2015role, snijder2020unifying}.

Further, similar to other RNA viruses, evidence suggests that coronaviruses interact with the cellular autophagy pathway to enhance virus replication \cite{miller2020coronavirus, snijder2020unifying}. The ER-derived double membraned vesicles in the host cytoplasm are similar to autophagosome. This similarity suggests that coronaviruses may mimic the cellular autophagy pathway for its replication.  Based on this brief survey, the task executed by nsp6 can be summarized as follows. Nsp6 maintains a balance in host autophagy signaling - (i) to escape viral degradation through lysosome-dependant mechanism, (ii) while availing the energy support of host nutrients for viral replication. However, the link between formation of DMVs, autophagy signaling and coronavirus replication has not been clearly established. Employing the CAML model, Section~\ref{CAML_nsp6_result} reports a comparative study of execution of the task by CoV-2 and CoV nsp6.

\subsection{Task executed by nsp8}
\label{Task_nsp8}

\vspace{2mm}
In association with nsp7, the nsp8 forms a complex of cylindrical structure that stabilizes the RdRp (RNA dependent RNA polymerase) function of nsp12. Consequently, a number of authors \cite{hillen2020structure, subissi2014one, biswal2021two, reshamwala2021mutations} have concentrated on analysis of the super-complex nsp12 - nsp8 - nsp7. The crystal structure reported in \cite{biswal2021two} provides the details of the oligomer interfaces of the nsps. All the authors have confirmed the fact that nsp7 and nsp8 complex impart processivity to nsp12. In Silico mutational study on nsp8 presented in \cite{reshamwala2021mutations} reports percentage interaction of the mutants on amino acids D50, K58, K82, S85, D99, P116, P183, R190 of nsp8. The mutants for which interaction with nsp12 gets significantly reduced are captured in the results reported in Section~\ref{CAML_nsp8_result} from mutational study reported out of CAML model.

\subsection{Task executed by nsp10}
\label{Task_nsp10}

\vspace{2mm}
A number of authors \cite{rogstam2020crystal, kasuga2021innate, bouvet2014coronavirus, saramago2021new, krafcikova2020structural, lin2021crystal} have reported structural details of nsp10 including its interaction with nsp14 and nsp16. The nsp10 has been identified as a central player in RNA replication due to its interaction/stimulation of the activities -  (i) ExoN domain of nsp14; and                                                                                                                                                                                                          (ii) the 2'-O-MTase activity of nsp16.
 
High sequence similarity of CoV-2 and CoV gets reflected in their structural similarity. There are only two amino acid changes between the two SARS strains - A23, and K113 for CoV-2, while CoV displays P23, and R113. Heterodimer structure has been reported for nsp10 in association with nsp14 or nsp16. The nsp10 secondary structure displays five $\alpha$-helices, a single 310-helix, and three $\beta$-strands.  It exhibits a fold structure around two zinc binding sites - (i) one formed by four AA (Cys74, Cys77, Cys90, His83) located between the helices $\alpha 2$ and $\alpha 3$; and (ii) second one formed by four cysteine residues (Cys117, Cys120, Cys128, Cys130). Zinc binding sites impart stability to nsp10 acting as cofactor for nsp 14 and nsp16 \cite{bouvet2014coronavirus}.  
Considering the AAs associated with Zn-binding sites, Section~\ref{CAML_nsp10_result} reports a comparative study of execution of CoV-2 and CoV nsp10 task with CAML model.

\subsection{Task executed by nsp12}
\label{Task_nsp12}

\vspace{2mm}
Analysis of nsp12 function and its structure in association with its cofactors nsp8, nsp7 are reported by many authors \cite{hillen2020structure, naydenova2021structure, kirchdoerfer2019structure, yan2021cryo, biswal2021two, maio2021fe, peng2020structural, aftab2020analysis, wang2021sars, gao2020structure}. The primary and secondary tasks of nsp12 can be noted as follows.\\
Primary Task - regulates replication and transcription of viral genome.\\                                                                                                                  Secondary Task - Nsp12 also inhibits IRF3-triggered IFN-$\beta$ production via a mechanism independent of its enzymatic activity.\\  
A replication-transcription complex (RTC) is formed by nsp12 in association with other nsps. The RTC carries out RNA synthesis, capping and proofreading.  The catalytic subunit of the nsp12 ligates two iron-sulphur metal clusters. These metal binding sites are essential for replication and interaction with the viral helicase; these iron-sulphur clusters can be also viewed as cofactors of nsp12. From survey of published literature, the motifs and important AAs are curated and listed in Table~\ref{task_nsp12} for three viruses having 932, 928, and 932 AAs for CoV-2, CoV, and MERS respectively.

\begin{table}[h]
\centering
\resizebox{\textwidth}{!}{ 
\begin{tabular}{|c|c|c|c|c|}
\hline 
Virus & Motif & Active Sites & Fe-S Catalytic site & Fe-S Interface  \\ 
\hline 
CoV-2 & L731, Y732, R733 & S759, D760, D761, K798, S814 & C487, H642, C645, C646 & H295, C301, C306, C310 \\ 
\hline 
CoV & V727, Y728, R729 & S755, D756, D757, K794, D810 & C483, H638, C641, C642 & H291, C297, C302, C306 \\ 
\hline 
MERS & L728, Y729, V730  & S760, D761, D762, K799, S815 & C488, H643, C646, C647 & H296, C302, C307, C311 \\ 
\hline 
\end{tabular} } 
\caption{The motif and catalytic sites of nsp12}
\label{task_nsp12}
\end{table}

Considering the associated AA noted above, Section~\ref{CAML_nsp12_result} reports a comparative study of execution of CoV-2 and CoV nsp12 task with CAML model.

\subsection{Task executed by nsp14}
\label{Task_nsp14}

\vspace{2mm}
The nsp14 is a bi-functional enzyme consisting of an exoribonuclease (ExoN) domain and a methyltransferase (MTase) domain. A number of authors \cite{tahir2021coronavirus, ogando2020enzymatic, kasuga2021innate, saramago2021new, lin2021crystal, ma2015structural, yuen2020sars, yoshimoto2021biochemical} have confirmed structural stability and increased efficiency for execution of nsp14 function in association with its cofactor nsp10. A mutation on nsp10 altering the binding ability of nsp10 to nsp14, reduces the capacity of nsp14 ExoN to efficiently cleave RNA. The second task of nsp14 has been shown to be independent of cofactor nsp10. The nsp14 catalyzes nucleoside monophosphate employing two divalent metal ions. The enzymatic activity of the nsp14 Exoribonuclease (coordinated by metal ions) are executed by the five catalytic (DEEDH) amino acids (D90, E92, E191, D273, H268). Inactivating the exoribonuclease (ExoN) activity has been shown to be lethal for SARS CoV-2 \cite{ogando2020enzymatic}. In summary, the nsp14 enzyme executes two interlinked tasks -\\                               
First task - It has a 3'to 5' exoribonuclease (ExoN) that mediates proof reading during viral replication. This activity is crucial for replication with high fidelity.\\                                                                                                                                                                                                                                            Second task- nsp14 covers a guanine-N7-Methyltransferase MTase that supports mRNA capping necessary for mRNA translation, while evading host immune response.  

On considering the AAs (DEEDH) at the catalytic site, Section~\ref{CAML_nsp14_result} reports a comparative study of execution of the tasks of CoV-2 and CoV nsp14 with CAML model.

\subsection{Task executed by nsp15}
\label{Task_nsp15}

\vspace{2mm}
The structural and functional aspects of nsp15 are reported in \cite{yoshimoto2021biochemical, pillon2021cryo, vijayan2021structure, deng2018old, kim2020crystal, zhang2018structural, guarino2005mutational, mandilara2021role, hackbart2020coronavirus, yuen2020sars}. The nsp15 is a uridine specific endoribonuclease (referred to as EndoU). It cleaves viral polyuridine RNA sequence that activates host immune sensors. Through the cleavage operation, nsp15 enzyme enables the virus to evade detection of virus by host defense system. The three domains of nsp15 are NTD (1 to 62), middle domain (63 to 191 covering flexible loop structure) and large CTD (192 to 345). From the structural analysis reported in \cite{yoshimoto2021biochemical, kim2020crystal, guarino2005mutational}, the nsp15 has been confirmed as a hexamer derived out of a dimer of trimers. The largest difference between SARS-CoV-2 and SARS-CoV seems to occur in the position of middle domains \cite{kim2020crystal}. This structural characteristic gets reflected in the results reported in Section~\ref{CAML_nsp15_result} with CAML model. Six important amino acids associated with EndoU catalytic activity has been confirmed as (H234, H249, K289 ,V294, Y342, Q346). The corresponding AA for MERS are (H231, H246, K286, V291, Y339, Q343). Based on the AAs associated with Catalytic function of nsp15, Section~\ref{CAML_nsp15_result} reports the results derived out of CAML model for CoV-2 and CoV nsp15.

\subsection{Task executed by nsp16 - 2'-O-methyltransferase}
\label{Task_nsp16}

\vspace{2mm}
A large number of authors \cite{almazan2006construction, chang2021nsp16, sk2020computational, krafcikova2020structural, aouadi2017binding, wang2015coronavirus, jiang2020repurposing} have dealt with different aspects of nsp16 in respect of virus RNA capping mechanism and the resulting down regulation of host immune response. Genetic degradation of SARS-CoV nsp16 significantly reduces the synthesis of viral RNA \cite{almazan2006construction}. Viruses, as surveyed in \cite{roossinck2017symbiosis}, evolves different strategies to survive in host cell. One such instance can be observed in nsp16 function.                                                   

Viruses replicating in cytoplasm have evolved 2'-O-methyltransferase (2'O-MTase) to autonomously modify their mRNAs. Such viral RNA modifications enable virus to avoid the host cell's discrimination between self and non-self mRNA. This characteristic allows viral mRNA to mimic that of human cell and is critically involved in subverting the induction of Interferon IFN1. Overcoming the host immune response is of paramount importance for the success of any viral infection in a host cell. The nsp16 is involved in viral RNA capping through methylation. With nsp10 as cofactor, nsp16 mediated 2'-O-methylation of coronavirus RNA prevents host recognition and decreases host immune response. The cap installation process is catalyzed by nsp10, as a zinc-binding protein that binds to viral RNA and stabilizes the binding pocket in NSP16, forming a stable complex. The binding site AA (K46, D130, K170, and E203), the KDKE motif of nsp16, are conserved among CoV-2, CoV, MERS. Section~\ref{CAML_nsp16_result} reports a comparative study of execution of CoV-2 and CoV nsp16 task with CAML model considering the associated amino acids at the catalytic sites.

In \cite{hazari2022analysis} we reported the CAML model results for Envelope Protein E - investigating its contribution towards higher transmissibility of Cov-2 compared to CoV. We hypothesized that difference of CAML model parameters between a pair of virus proteins of similar length and executing same function, model the difference in their structure - function. The same analogy holds good for a mutant and its wild counterpart. The hypothesis has been validated in \cite{hazari2022analysis} from mutational study on three diverse viruses - (i) human HBB hemoglobin protein associated with sickle cell anemia, (ii) mutants reported in envelope protein of covid-2 infected patients, and (iii) Hepatitis B Virus (HbV) x protein. For each of these case studies, the CAML model parameter differ between wild and specific mutant that represent deviation of the structure-function of the mutant from that of wild leading deviation/disruption of normal function.

\section{Cellular Automata (CA) Preliminaries}
\label{CA_preli} 

CA is a dynamical system discrete in space and time. The space is represented by a regular lattice in one, two or higher dimensions. Each site on the lattice also referred to as a cell, can be in one of a finite number of states. The next state of a cell depends on the current state of its neighbours and the associated next state function. Neumann \cite{von1996theory} proposed the CA model for constructing self-replicating machine employing 2000 cells, each holding 29 states. Subsequently, Wolfram \cite{wolfram1983statistical} proposed a simpler version of CA with two states per cell and three-neighbourhood. Research of a large number of authors from diverse disciplines has enriched the field of CA \cite{codd1968cellular, conway1970game, wolfram2002new, chaudhuri2018new, ppc1}. The book \cite{ppc1} covers a comprehensive survey of CA theory and wide varieties of applications. Further, the universal appeal of CA model based on local interaction can be ascertained from the materials presented in the book \cite{ridley2015evolution} that highlights the effect of local (temporal and physical) changes transforming the human society from pre-historic to modern age.

\begin{figure}[h]
\begin{center}
     \includegraphics[scale=0.55]{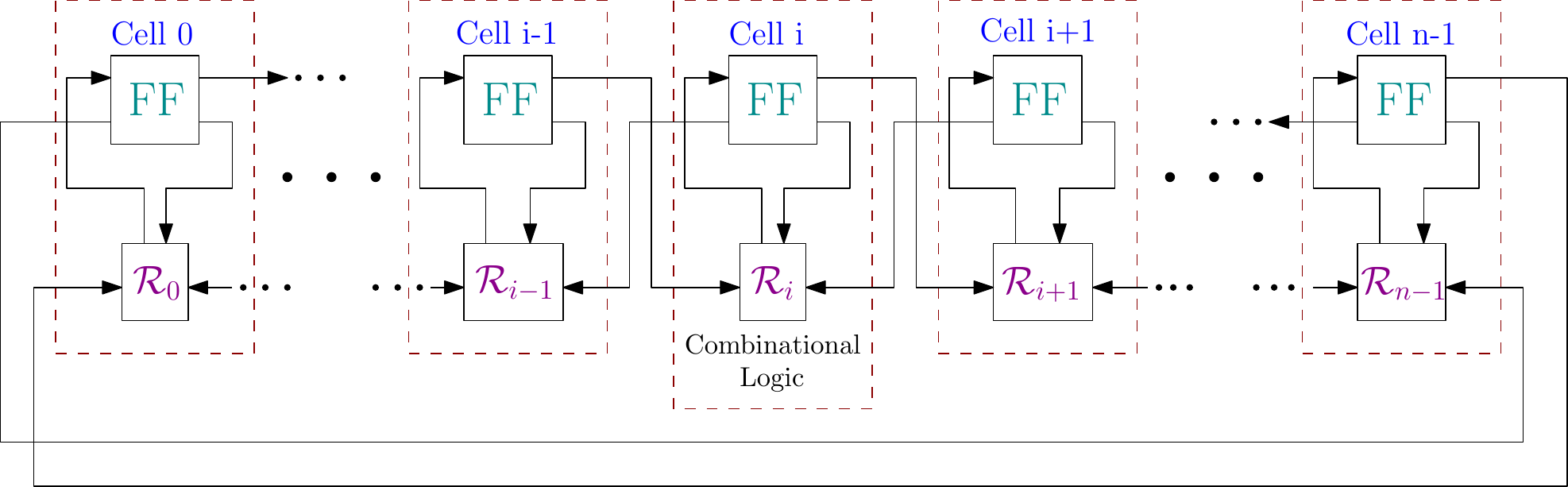}
     \caption{Implementation of an $n$-cell 3-neighborhood CA under periodic boundary condition}
     \label{Hardware-ECA-Periodic}
\end{center}
\end{figure}

\vspace{-3mm}
The CA model reported in this paper employs simplest CA structure with periodic boundary, three neighborhood, and two state per cell - referred to as 3NCA in the rest of the paper. The implementation of an $n$-cell 3-neighbourhood CA under periodic boundary condition is shown in Figure~\ref{Hardware-ECA-Periodic}. The middle cell is marked as $i^{th}$ cell while its left and right neighbors are denoted as $(i - 1)^{th}$ and $(i + 1)^{th}$ cell respectively. Consequently, $8(2^3)$ combinations exist for the triplet of current state values $<a_{i-1}  a_i  a_{i+1}>$ of $(i - 1)^{th}$ , $i^{th}$ and $(i + 1)^{th}$ cells. The next state function of a cell is shown as a combinational logic in Figure~\ref{Hardware-ECA-Periodic}. The decimal value of the 8 bits of cell next state is referred to as `Rule' of evolution of a CA cell. There are $2^8 = 256$ such rules for three-neighborhood CA. The decimal value derived out of 8 bit string of two CA rules (45) and (105) are illustrated in Table~\ref{Trules} Row 4 and 5. Eight different input combinations of a CA rule is a triplet of binary bits - 111(7), 110(6), 101(5), 100(4), 011(3), 010(2), 001(1), 000(0), are noted on row 1 of Table~\ref{Trules}. Each combination represents the current state of $(i - 1)^{th}$, $i^{th}$, and $(i + 1)^{th}$ cells which can be viewed as three binary variables. In subsequent discussions we refer to the combinations of three input variables as Rule Min Terms (RMTs).  A RMT is referred to by its decimal value that varies from 0 to 7. Row 3 of Table~\ref{Trules} shows the cell next as $b_k$ ($b_k$ = 0 or 1), for $k$= 0 to 7. By convention, a CA rule is expressed as decimal counterpart of binary bit string $<b_7b_6b_5b_4b_3b_2 b_1b_0>$. So conventional weight assignment follows for conversion of 8-bit string of a rule to its decimal value with $w_7 = 2^7$, $w_6 = 2^6$ and so on. The decimal rule number 45 (00101101) is derived from its binary bit string as - $\sum_{k=0}^7 w_k b_k = 2^5 + 2^3 + 2^2 + 2^0 = 32 + 8 + 4 + 1 = 45$. A CA rule can be viewed as a Transform that accepts 3 bit input $k = <a_{i-1} a_i a_{i+1}>$ and generates single bit output $b_k$ in the next time of evolution of the cell. The cells of a 3NCA employs one of the 256 rules (0 to 255).

{\small
\begin{table*}[h]
\[
\begin{array}{|cccccccccc|}
\hline
{\rm Present~ State:} &  111 & 110 & 101 & 100 & 011 &  010 &  001 &  000 & Rule \\ 
(RMT)	& (7) & (6) & (5) & (4) & (3) & (2) & (1) & (0) & \\ 
{\rm  Next~ State ~(b_k)}    & b_7  & b_6 & b_5 & b_4 & b_3  & b_2  & b_1  & b_0  &  \\ \hline
  {\rm ~ (i)~ Next~ State:}    &   0  &  0 &  1  &  0  &   1  & 1  &   0  &   1  & 45 \\
{\rm (ii)~ Next~ State:}    & 0  &  1 &  1  &  0  &   1  & 0  &   0  &   1   & 105 \\
\hline
\end{array}
\]
\vspace{-3mm}
\caption{Look-up table for rule 45 and 105}
\label{Trules}
\end{table*}
}

In order to design rules to build CA model for a physical system, it is convenient to use the following format for a rule.  We employ this format for design of rules for 20 amino acids in next section.

\vspace{2mm}
\noindent
\textbf{1-major and 0-major (7653 4210) Format:} Out of 8 RMTs, the binary string of each of the 4 RMTs (7, 6, 5, 3) has two 1's, while RMTs (4, 2, 1, 0) has two 0's - these two classes are  referred to as 1-Major  and 0-Major RMTs respectively in Table~\ref{ECA_rule_Table} that shows rules 45 and 105 in this format.

\begin{table}[h]
\centering
\resizebox{0.95\textwidth}{!}{
\begin{tabular}{|c|cccc|cccc|}
\hline 
\multirow{2}{*}{Rule} & \multicolumn{4}{c|}{1-Major RMTs} & \multicolumn{4}{c|}{0-Major RMTs} \\ 
 & 7(111) & 6(110) & 5(101) & 3(011) & 4(100) & 2(010) & 1(001) & 0(000) \\ 
\hline 
45 & 0 & 0 & 1 & 1 & 0 & 1 & 0 & 1 \\ 
\hline 
105 & 0 & 1 & 1 & 1 & 0 & 0 & 0 & 1 \\ 
\hline 
\end{tabular} }
\caption{Representation of two CA rules 45 and 105 in 1-major and 0-major format}
\label{ECA_rule_Table}
\end{table}

\begin{figure}[h]
\hfill
\subfigure[Ten cells  of a CA configured with ten rules \label{ECARules}]{\includegraphics[width=5cm, height=8mm]{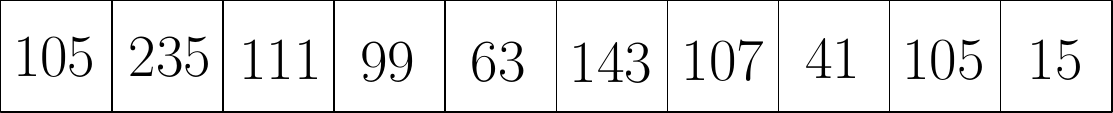}}
\hfill
\subfigure[Rule Vector\label{Rule-Vector}]{\includegraphics[width=5cm, height=8mm]{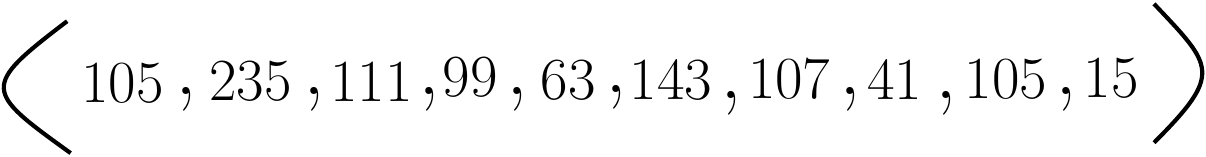}}
\hfill
\caption{Rule vector for a 10 cell 3NCA}
\end{figure}

\section{CA Rule Vector for Amino Acids of a Protein Chain}
\label{Sec:CA_rule_for_AA}

Design of CA model for a physical system involves design of rule for each CA cell. A cell models the smallest component of the system so that the physical domain features of the component get represented in the binary bit pattern of its rule. Once the rule design step is complete, we can represent a 3NCA as a Rule Vector - a string of 3NCA rules (0 to 255). Figure~\ref{Rule-Vector} shows the rule vector for a 10 cell CA with the rules specified in Figure~\ref{ECARules}. The CA evolves at each time step through local interaction of each cell with its neighborhood. The design proceeds through a number of iterations to build a model that characterizes the system with efficient mapping of model parameters derived out of CA evolution to the physical domain features of the system. The string of 3NCA Rule Vector can be employed to model the biological strings - DNA, RNA, Codon string, and Amino Acid chain of a protein.

\begin{figure}[h]
\begin{center}
\includegraphics[width=12cm, height=8.5cm]{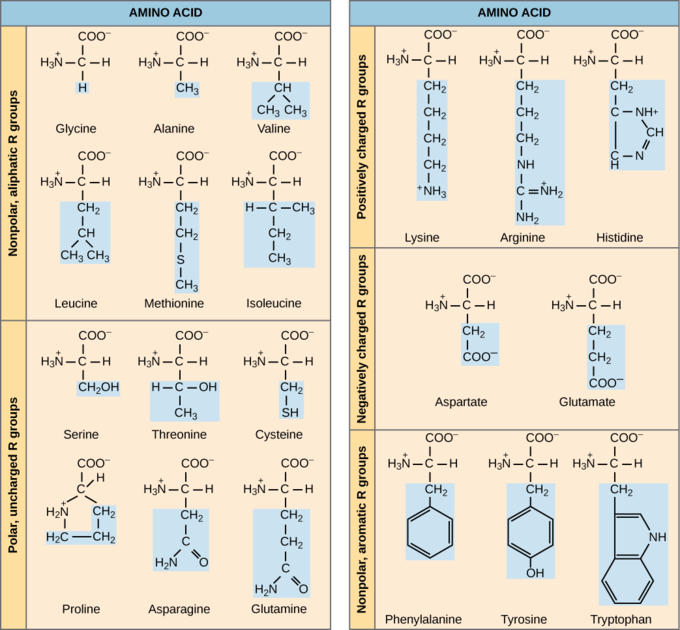}
\caption{Atomic structure of 20 common amino acids(AA) divided into 5 groups}
\label{20_AAcide}
\end{center}
\end{figure}

The CA model for 20 amino acids has been derived from the first principle by considering the atomic structure of amino acid molecules. Figure~\ref{20_AAcide} shows atomic structure of 20 amino acids divided into 5 different groups based on the property of the side chain referred to as Residue (R). The side chain property - non-polar, polar, positively charge, negatively charged, aromatic - is defined by the structure of interconnected atoms. For design of CA rules, the atoms are divided into two groups -  H-atom and non-H atom (Carbon C, Nitrogen N, Oxygen O, Sulphur S), each having different proton count in nucleus and electron count in outer shells. The CA model for hydrogen atom H, lightest one in the periodic table elements, treated differently from non-H atoms.

\subsection{Design of CA rule for Amino Acid (AA) Chain of Protein}

\vspace{2mm}
Each amino acid has a common Back-Bone with 9 atoms - 5 non-H atoms (2 C, 2 O, 1 N) and 4 H atoms. The side chain residue (R) has widely different number of atoms - 1 atom (for Gly - G) to 18 atoms (for Arg - R). CA rule for amino acid is designed as a composite module of backbone connected to side chain. This is a departure from the convention of referring to an amino acid as a residue. In the rest of the paper, we refer to an amino acid (AA) as a composite molecule covering both backbone and residue.

\subsubsection{Design Methodology}
\label{Design_meth_AA}

\vspace{2mm}
3NCA rules for amino acid molecules are defined as a sequence of rules (decimal value 0 to 255). On considering the atomic structure of amino acids, it is convenient to design the rules by considering the RMT string <7 6 5 4 3 2 1 0> of a rule in the form of 1-Major and 0-Major RMTs as <7653   4210>(see Table~\ref{ECA_rule_Table}). The H atoms are assigned as `1' in the next state of 0-major RMTs, while non-H atoms are assigned as `1' in the next state of 1-major RMTs. However, we do not populate more than 6 number of `1's in the 8-bit pattern of a rule. Hence, rather than a single rule, the design employs a string of rules if the molecule has more than 4 H atoms or non-H atoms.

\subsubsection{Design of CA rule for Amino Acid Backbone}

\vspace{2mm}
As per this design methodology, a string of two CA rules designed for 9 atom amino acid  backbone as <107 99>is reported in Table~\ref{AA_backbone}, where the first column shows design of two rules expressed in the 1-Major 0-Major format for 9 atoms (5 non-H atoms followed by 4 H-atoms). The fourth column presents two rules in the <7 6 5 4 3 2 1 0> format; finally, column 5 shows the rule string with rules expressed in decimal value. Thus, the CA backbone is modelled with two CA cells configured with a string of two CA rules <107 99>.

\begin{table}[h]
\begin{center}
\resizebox{\textwidth}{!}{ 
\renewcommand{\arraystretch}{1.5} 
\begin{tabular}{|c|c|c|c|c|}
\hline 
Rules for AA backbone in & Number  & Number of   & Rules for AA Backbone in & Rule string for \\  
<7653 4210> format & of & atoms in (AA)  & <7 6 5 4 3 2 1 0> format &  AA backbone \\                                                                                                                                                                                                                                                             
                   &   rules    & backbone         &                      & in Decimal  \\             
\hline 
<0111 0011><0110 0011> & 2 & 2C-N-2O-4H(9) & <0110 1011><0110 0011> & <107 99> \\ 
\hline 
\end{tabular} }
\caption{Design of CA rule string for amino acid backbone having 9 atoms - two carbon, one nitrogen, two oxygen, and 4 hydrogen atoms}
\label{AA_backbone}
\end{center}
\end{table}

\vspace{-8mm}
\subsubsection{Design of CA rule for Amino Acid (AA)}
\label{CA_rule_forAA}

\vspace{2mm}
The atomic structure of 20 amino acids are presented in Figure~\ref{20_AAcide}. Following the design methodology reported in Section~\ref{Design_meth_AA}, rule strings for 20 AA are designed and reported in Table~\ref{AA_Composite_rule_Vector}. The first column presents the design of rule string expressed in 1-major 0-major <7653 4210>format with column 2, 3, 4 respectively showing - number of rules, AA name, number of atoms (H and non-H) in the side chain. The rule string is shown in column 5 in the <7 6 5 4 3 2 1 0> format  and the corresponding decimal value (0 to 255)  in column 6. Finally, the composite rules for an amino acid with backbone and side chain rule concatenated is reported in column 7 of Table~\ref{AA_Composite_rule_Vector}. For example, the composite Rule string <107 99><63 11 15> in column 7 is designed for the amino acid Ile (I) with 9 atoms for common backbone and 13 atoms for side chain shown in $6^{th}$ row proceeds as follows. First a string of two rules <107 99> is designed for backbone (Table~\ref{AA_backbone}). A string of 3 rules is next designed for 13 atoms of side chain with 9 H atoms preceded by 4 non-H atoms. Intermediate columns 5 and 6 show the rules expressed in RMT format <7 6 5 4 3 2 1 0> and their decimal counterpart. Last four rows of the Table~\ref{AA_Composite_rule_Vector} display the nucleotide base triplet for 20 amino acids and 3 stop codons. In the rest of the paper, an amino acid is referred to as AA modeled with 3 to 5 cells configured with a string of 3 to 5 rules. The AAs are serially noted with met (m) as the first  AA of an amino acid chain of a protein.

\begin{table}[h]
\centering
\resizebox{1.0\textwidth}{!}{ 
\renewcommand{\arraystretch}{1.35}
{\bf
\begin{tabular}{|c|c|c|c|c|c|c|}
\hline 
Rules for AA side chain in  & No. of  & Amino acid & No. of & Rules for AA side chain in & Rule string for AA  & Composite Rule String  \\ 
<7653 4210>format & rules & (AA) name & atoms & <7 6 5 4 3 2 1 0> format & side chain in Decimal & for Amino acid \\ 
\hline 
<0000 0001> & 1 & Gly(G) & H(1) & <0000 0001> & <1> & <107 99><1> \\ 
\hline 
<0001 0111> & 1 & Ala(A) & CH3(4) & <0000 1111> & <15> & <107 99><15> \\ 
\hline 
<0011 1111><0001 0111> & 2 & Val(V) & C3H7(10) & <0011 1111><0000 1111> & <59 15> & <107 99><59 15> \\ 
\hline 
<0001 0011><0001 0111> & 3 & Leu(L) & C4H9(13) & <0000 1011><0000 1111> & <11 15 63> & <107 99><11 15 63> \\ 
<0011 1111> &   &  &  & <0011 1111> &  &  \\
\hline 
<0011 1111><0011 0111> & 2 & Met(M) & SC3H7(11) & <0011 1111><0011 0111> & <63 47> & <107 99><63 47> \\ 
\hline 
<0011 1111><0001 0011> & 3 & Ile(I) & C4H9(13) & <0011 1111><0000 1011> & <63 11 15> & <107 99><63 11 15> \\
 <0001 0111> &   &  &  & <0000 1111> &  &  \\ 
\hline 
<0011 0111> & 1 & Ser(S) & COH3(5) & <0010 1111> & <47> & <107 99><47> \\ 
\hline 
<0011 0011><0001 0111> & 2 & Thr(T) & C2OH5(8) & <0010 1011><0000 1111> & <43 15> & <107 99><43 15> \\ 
\hline 
<1001 0111> & 1 & Cys(C) & SCH3(5) & <1000 1111> & <143> & <107 99><143> \\ 
\hline 
<0001 0011><0011 1111> & 2 & Pro(P) & C3H6(9) & <0000 1011><0011 1111> & <11 63> & <107 99><11 63> \\ 
\hline 
<0001 0011><0111 0011> & 2 & Asn(N) & C2NOH4(8) & <0000 1011><0110 1011> & <11 107> & <107 99><11 107> \\ 
\hline 
<0011 1111><0111 0011> & 2 & Gln(Q) & C3NOH6(11) & <0011 1111><0110 1011> & <63 107> & <107 99><63 107> \\ 
\hline 
<0011 1111><0011 1111> & 3 & Lys(K) & C4NH11(16) & <0011 1111><0011 1111> & <63 63 15> & <107 99><63 63 15> \\ 
 <0001 0111> &   &  &  & <0000 1111> &  &  \\
\hline 
<0011 1111><0111 0111> & 3 & Arg(R) & C4N3H11(18) & <0011 1111><0110 1111> & <63 111 63> & <107 99><63 111 63> \\ 
 <0011 1111> &   &  &  & <0011 1111> &  &  \\
\hline 
<0111 0111><0111 0011> & 2 & His(H) & C4N2H5(11) & <0110 1111><0110 1011> & <111 107> & <107 99><111 107> \\ 
\hline 
<1111 0011> & 1 & Asp(D) & C2O2H2(6) & <1110 1011> & <235> & <107 99><235> \\ 
\hline 
<0001 0011><1111 0011> & 2 & Glu(E) & C3O2H4(9) & <0000 1011><1110 1011> & <11 235> & <107 99><11 235> \\ 
\hline 
<0001 0011><0111 0011> & 3 & Phe(F) & C7H7(14) & <0000 1011><0110 1111> & <11 107 111> & <107 99><11 107 111> \\ 
 <0111 0111> &   &  &  & <0110 1111> &  &  \\
\hline 
<0001 0011><0111 0111> & 3 & Tyr(Y) & C7OH7(15) & <0000 1011><0110 1111> & <11 111 111> & <107 99><11 111 111> \\ 
 <0111 0111> &   &  &  & <0110 1111> &  &  \\
\hline 
<0111 0111><1111 0011> & 3 & Trp(W) & C9NH8(18) & <0110 1111><1110 1011> & <111 239 111> & <107 99><111 239 111> \\ 
 <0111 0111> &   &  &  & <0110 1111> &  &  \\
\hline 

\end{tabular} } }

\begin{flushleft}
\small{\textbf{NonPolar Chain:} Gly (gg - c/g/a/t), Ala (gc - c/g/a/t), Val (gu - c/g/a/t), Leu (tt - a/g \& ct - c/g/a/t), Met (at - g), Ile (at - c/a/t);}\\

\small{\textbf{Polar uncharged:} Ser (ag - c/t \& tc - c/g/a/t), Thr (ac - c/g/a/t), Cys (tg - c/t), Pro (cc - c/g/a/t), Asn (aa - c/t), Gln (CA - G/A);}\\

\small{\textbf{Positively Charged: }Lys (AA - G/A), Arg (ag - g/a \& cg - c/g/a/t), His (ca - c/t);}\\
\small{\textbf{Negatively Charged:} Asp (ga - c/t),  Glu (ga - g/a);}\\

\small{\textbf{NonPolar Ring:} Phe (tt - c/t), Tyr (ta -  c/t), Trp (tg - g);}  

\small{\textbf{Stop codons:}  ( ta - g/a \&  tg - a)}
\end{flushleft}
 
\caption{CA rule string for AA residue (R) \& Concatenated Composite Rule String (backbone \& R) for AA}
\label{AA_Composite_rule_Vector}
\end{table}

\subsubsection{CA Rule Vector for AA Chain}

\vspace{2mm}
On completion of rule string design for each amino acid, rule vector can be derived for the AA chain of a protein by concatenating rule strings of each AA. Table~\ref{AA_Chain} shows nsp1 (non-structural protein 1) AA string for - (a) CoV-2 (2019) - 180 AA, (b) CoV (2003) - 180 AA, and (c) MERS (2012) - 193; partial Rule Vectors for three AA chains are illustrated.

\begin{table}
\centering
\resizebox{1.0\textwidth}{!}{ 
\renewcommand{\arraystretch}{1.2}
\begin{tabular}{|c|l|}
\hline 
\multicolumn{2}{|c|}{\textbf{ CoV-2 nsp1 AA sequence - 180 AA}} \\ 
\hline 
\multirow{3}{*}{AA Chain} & meslvpgfnekthvqlslpvlqvrdvlvrgfgdsveevlsearqhlkdgtcglvevekgvlpqleqpyvfikrsda \\ 
   & rtaphghvmvelvaelegiqygrsgetlgvlvphvgeipvayrkvllrkngnkgagghsygadlksfdlgdelgt  \\
   & dpyedfqenwntkhssgvtrelmrelngg   \\
\hline 
Partial & < <107, 99, 11, 235>, <107, 99, 47>, <107, 99, 11, 15, 63>, <107, 99, 59, 15>,  \\  
rule vector & <107, 99, 11, 63>, <107, 99, 1>, <107, 99, 11, 107, 111>, <107, 99, 11, 107>,  \\ 
            & <107, 99, 11, 235>, <107, 99, 63, 63, 15>, $\cdots$ > > \\
\hline 
\multicolumn{2}{|c|}{\textbf{CoV nsp1 AA sequence - 180 AA}} \\  
\hline 
\multirow{3}{*}{AA Chain} & meslvlgvnekthvqlslpvlqvrdvlvrgfgdsveealsearehlkngtcglvelekgvlpqleqpyvfikrsdal \\ 
   & stnhghkvvelvaemdgiqygrsgitlgvlvphvgetpiayrnvllrkngnkgagghsygidlksydlgdelgt \\
   & dpiedyeqnwntkhgsgalreltrelngg  \\
\hline 
Partial & < <107, 99, 11, 235>, <107, 99, 47>, <107, 99, 11, 15, 63>, <107, 99, 59, 15>, \\ 
rule vector & <107, 99, 11, 15, 63>, <107, 99, 1>, <107, 99, 59, 15>, <107, 99, 11, 107>, \\
            & <107, 99, 11, 235>, <107, 99, 63, 63, 15>, $\cdots$ > > \\
\hline 
\multicolumn{2}{|c|}{\textbf{MERS nsp1 AA sequence - 193 AA}} \\  
\hline 
\multirow{3}{*}{AA Chain} & msfvagvtaqgargtyraalnsekhqdhvsltvplcgsgnlveklspwfmdgenayevvkamllkkepllyvpir \\  
 & laghtrhlpgprvylverliacenpfmvnqlaysssangslvgttlqgkpigmffpydielvtgkqnillrkygrggy \\
 & hytpfhyerdntscpewmddfeadpkgkyaqnllkkligg \\ 
\hline 
Partial & < <107, 99, 47>, <107, 99, 11, 107, 111>, <107, 99, 59, 15>, <107, 99, 15>, \\ 
rule vector & <107, 99, 1>, <107, 99, 59, 15>, <107, 99, 43, 15>, <107, 99, 15>,  \\ 
            & <107, 99, 63, 107>, <107, 99, 1>, $\cdots$ > > \\
\hline 
\end{tabular} }

\caption{AA chain for nsp1 of SARS covid (CoV-2, CoV - each of length 180 AA), MERS covid (193 AA) retrieved from NCBI; partial Rule Vector (RV) for first ten AA (excluding Met (m) encoding start codon) reported following AA chain for three viruses}
\label{AA_Chain}
\end{table}

Once the CA Rule Vector is designed for a protein chain, we study the evolution of the CA generating a Signal Graph referred to as Cycle Length Signal Graph (CL Signal Graph) detailed next in Section~\ref{CA_Evolution_CL_Graph}.

\section{CA Evolution Generating Cycle Length Signal Graph (CL Signal Graph) for AA Chain of Protein}
\label{CA_Evolution_CL_Graph}

The CA model employs periodic boundary 3NCA (Figure~\ref{Hardware-ECA-Periodic}) for an amino acid chain of a protein excluding first amino acid Met of the chain. The cells of an $n$ cell CA are serially marked as $0, 1, 2, \cdots (n-1)$ , where $0^{th}$ cell and $(n-1)^{th}$ are neighbors of each other. An amino acid (backbone and side chain) is modelled with a string of 3 to 5 cells configured with 3 to 5 rules, as reported in Table~\ref{AA_Composite_rule_Vector}.  The CA evolves in successive time steps as per the Rule Vector that specifies the rule assigned for each CA cell. The cells of the CA are initialized with an alternating sequence of 1 and 0. The evolution continues till each cell settles down in a cycle, where the cell cycles through a set of binary states; the length of the cycle is referred to as Cycle Length (CL). The CL signal graph refers to the graph showing CL values for each cell on y-axis, with x-axis displaying serial location of CA cells. If a few cells do not settle down in a cycle even after 4000-time steps, these are marked with value CL(-5). (Note - The following fact has been experimentally verified through evolution of CoV-2, CoV, and MERS proteins: running the CA modeling protein chains for 4000 time steps generate minimal number of cells marked CL(-5), which do not reach a cycle; exponential decrease of count of CL(-5) cells has been observed with further increase of run time steps beyond 4000).

\subsection{CL Signal Graph Analytics for Wild Type AA Chain and Mutant}

\begin{figure}[h!]
\begin{center}
\includegraphics[width=13.3cm, height=4.9cm]{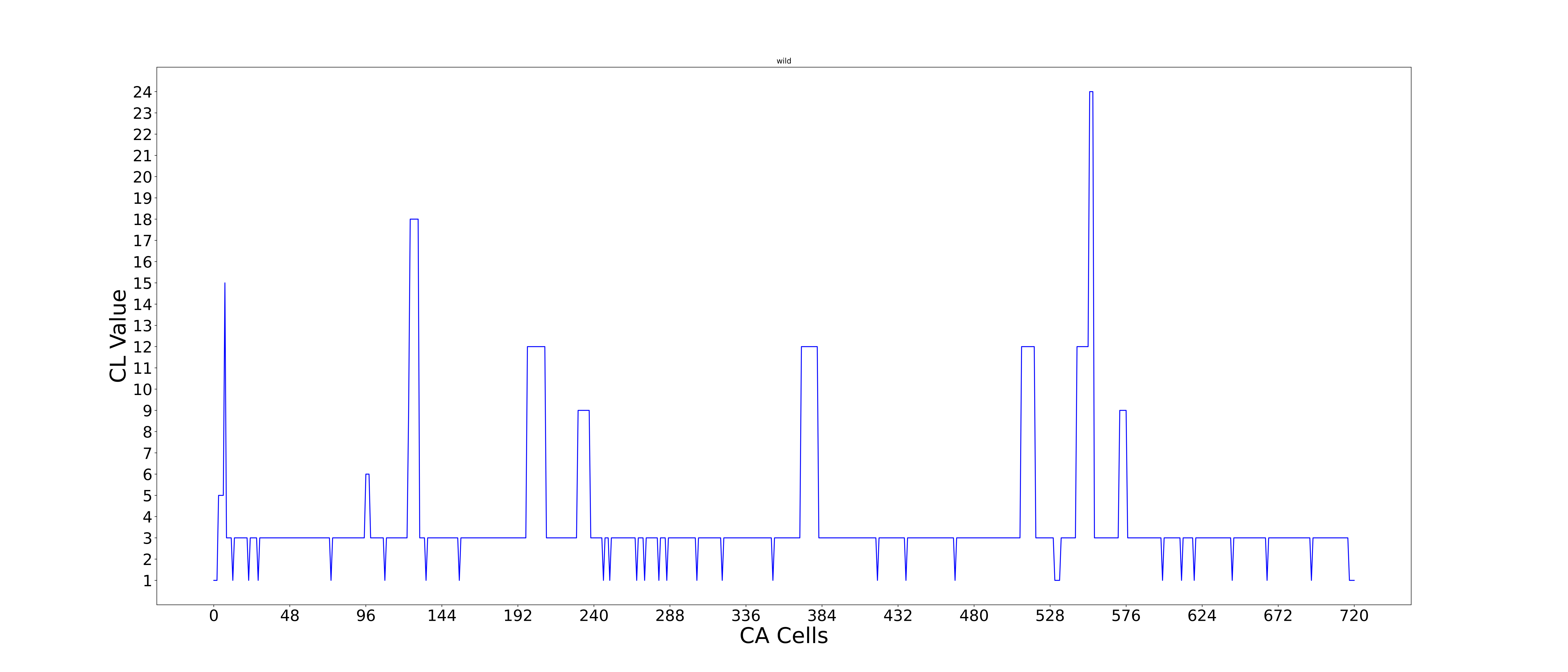}
\caption{CL signal graph for Cov-2 wild nsp1 AA chain}
\label{CL_Graph_Wild_AA_Cov2_nsp1}
\end{center}
\end{figure}

\begin{figure}[h!]
\begin{center}
\includegraphics[width=13.3cm, height=4.8cm]{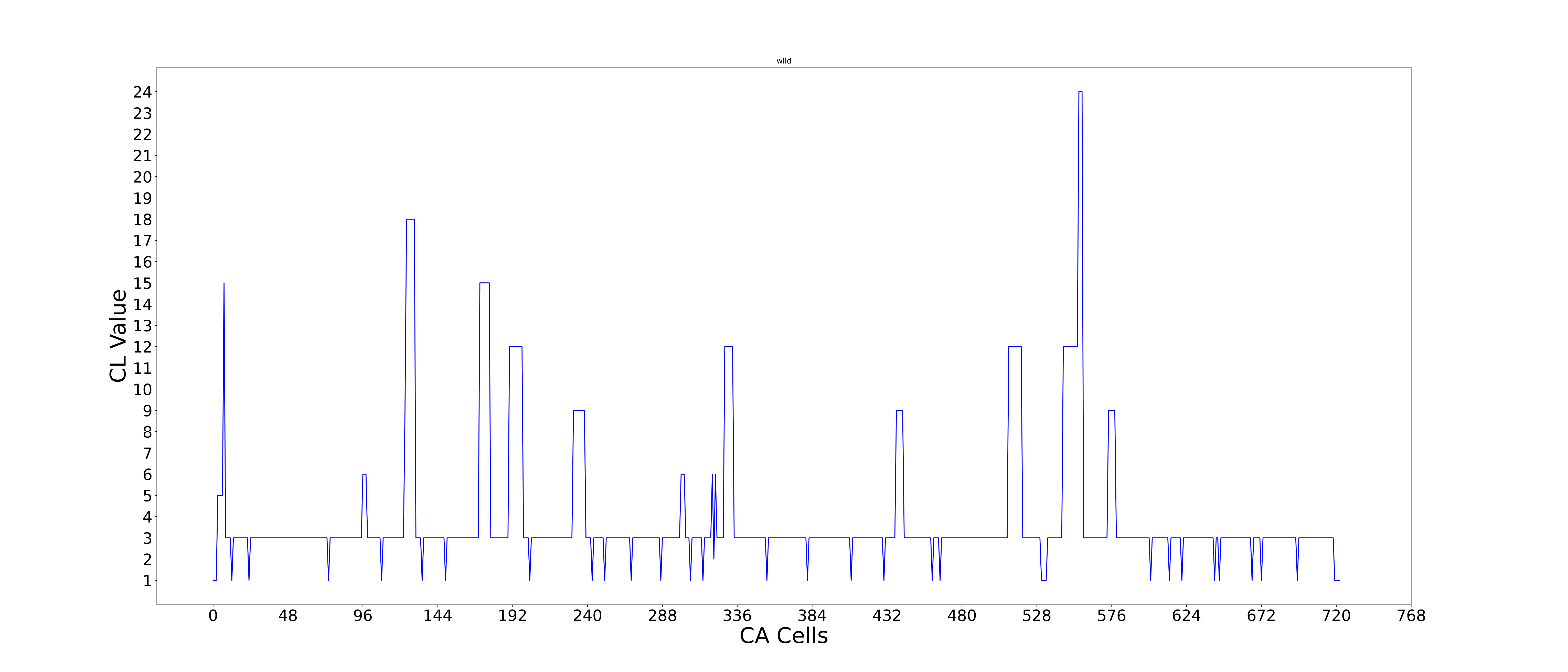}
\caption{CL signal graph for Cov wild nsp1 AA chain}
\label{CL_Graph_Wild_AA_Cov_nsp1}
\end{center}
\end{figure}

\begin{figure}[h!]
\begin{center}
\includegraphics[width=13.3cm, height=4.8cm]{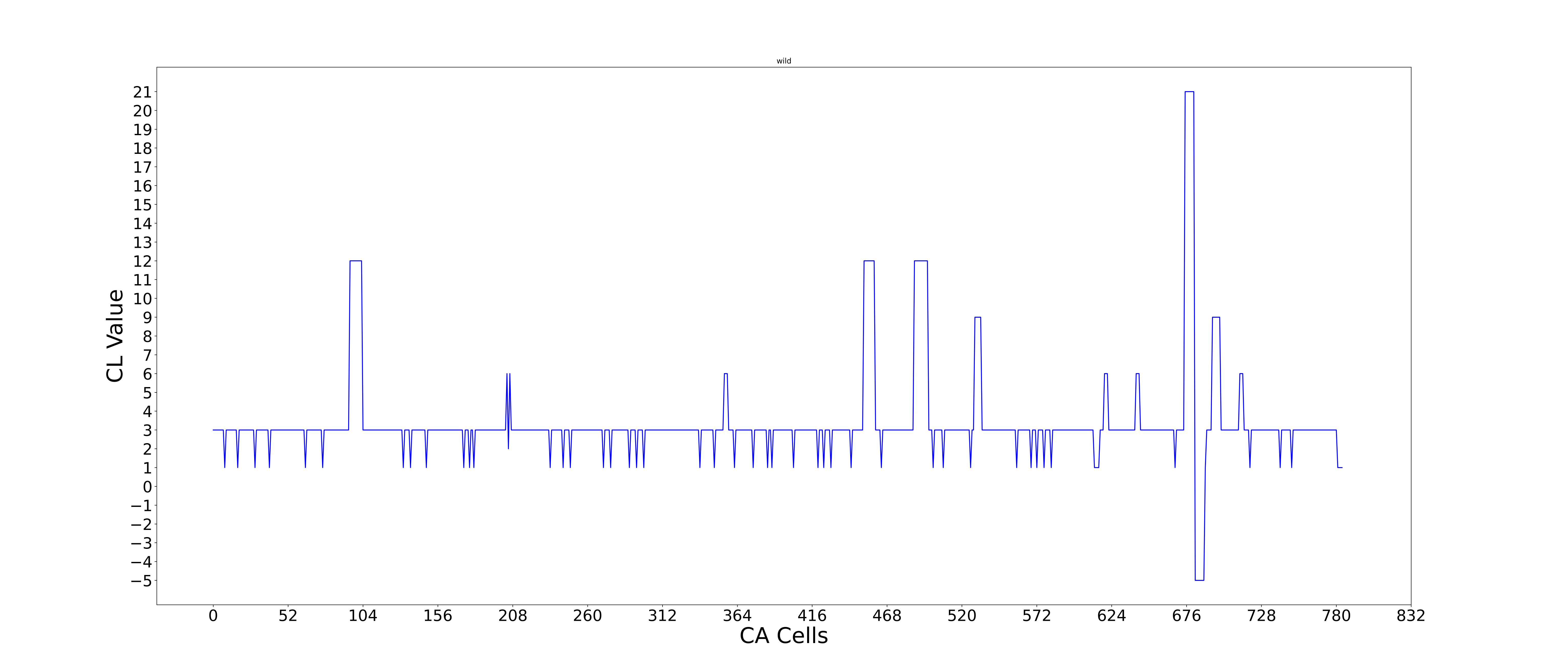}
\caption{CL signal graph for MERS wild nsp1 AA chain}
\label{CL_Graph_Wild_AA_MERS_nsp1}
\end{center}
\end{figure}

\vspace{2mm}
For the CAML model, signal graph analytics enable extraction of meaningful information from a graph relevant for mutational study of protein. Wild type AA chain refers to the one that can be observed in nature without any mutation. The chain with a mutation is termed as Mutant. CL signal graphs for wild type AA chain are shown in Figure~\ref{CL_Graph_Wild_AA_Cov2_nsp1}, \ref{CL_Graph_Wild_AA_Cov_nsp1}, \ref{CL_Graph_Wild_AA_MERS_nsp1}. In view of common backbone, the base line of the CL signal graph has the CL value 3.                                                                                                                                                                                                                   Signal graph analytics proceeds on identifying mcl (Maximum Cycle Length) and N-mcl (Next to mcl) signals. The mcl signal is the one that covers a number of cells with maximum cycle length value. There may be one or more mcl in a signal graph of an AA chain. An N-mcl signal has highest CL value less than mcl but greater than CL value 5. Such an evaluation N-mcl signal enables avoidance of noisy signals below the CL value 6 in a CL signal graph.

\subsubsection{CL Signal Graph for Wild Type AA Chain}

\vspace{2mm}
Figure~\ref{CL_Graph_Wild_AA_Cov2_nsp1}, \ref{CL_Graph_Wild_AA_Cov_nsp1},  \ref{CL_Graph_Wild_AA_MERS_nsp1} report the CL signal graphs  for evolution of CA as per Rule Vector for wild type AA chain noted in Table~\ref{AA_Chain} for CoV-2, CoV, and MERS nsp1.

\subsubsection{CL Signal Graph for Mutant AA Chain}

\vspace{2mm}
Ala mutagenesis, usually undertaken to determine the contribution of a specific AA in the function of a given protein. Instead of Ala, CAML model employs Cys mutagenesis to expose the contribution of the AA for protein function. The Property P5 presented in Section~\ref{ML_Fram_design_Method} reports the reason of such deviation from Ala to Cys.

\begin{figure}[h!]
\begin{center}
\includegraphics[width=13.3cm, height=4.7cm]{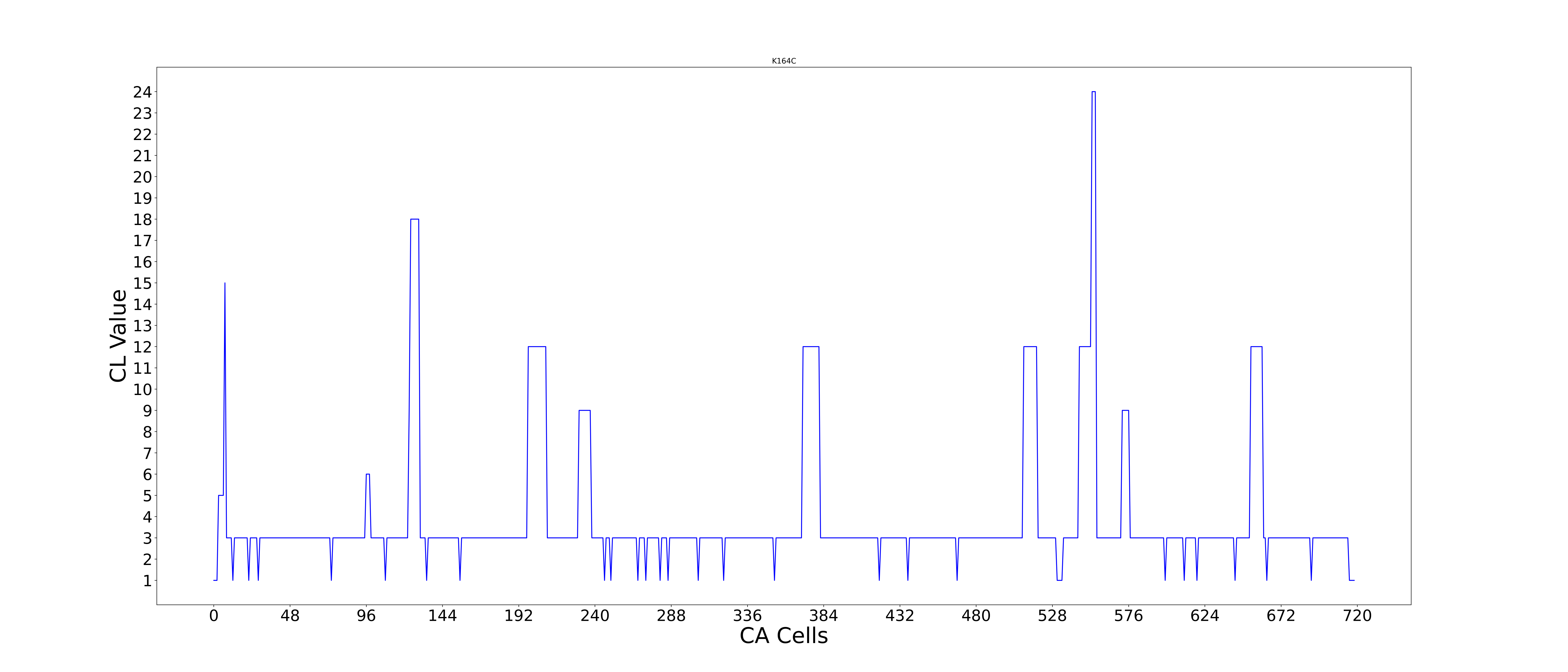}
\caption{CL Signal graph for mutant K164C for CoV-2 nsp1}
\label{CL_Graph_Mutant_K164C_Cov2_nsp1}
\end{center}
\end{figure}

\begin{figure}[h!]
\begin{center}
\includegraphics[width=13.3cm, height=4.7cm]{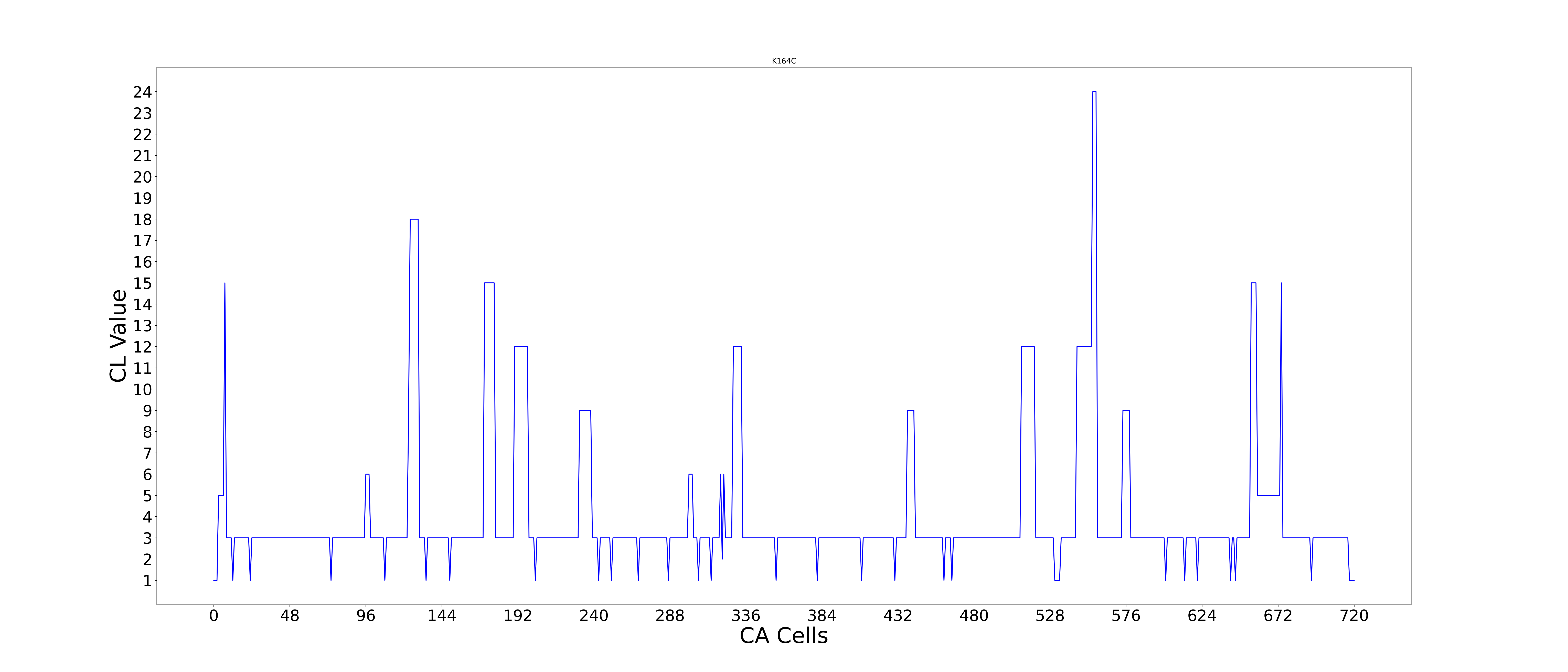}
\caption{CL Signal graph for mutant K164C for CoV nsp1}
\label{CL_Graph_Mutant_K164C_Cov_nsp1}
\end{center}
\end{figure}

\begin{figure}[h!]
\begin{center}
\includegraphics[width=13.3cm, height=4.7cm]{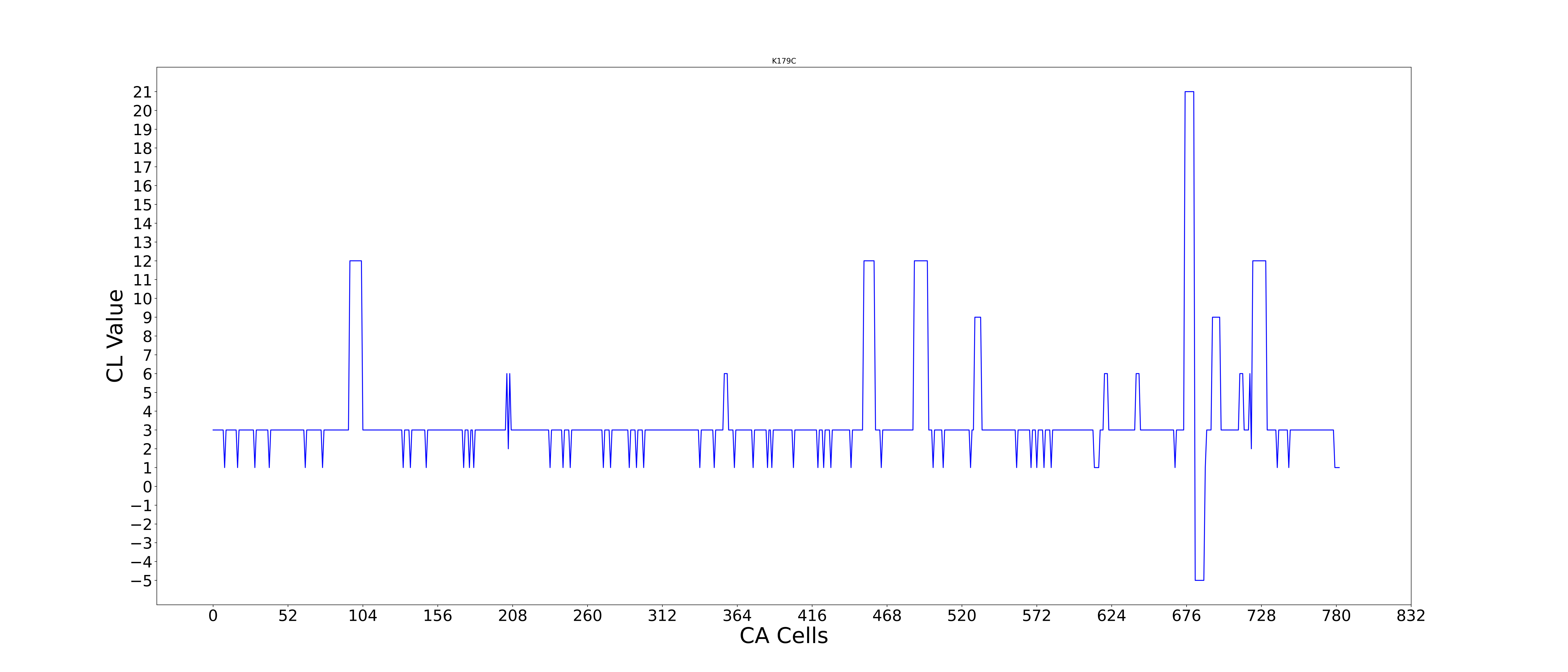}
\caption{CL Signal graph for mutant K179C for MERS nsp1}
\label{CL_Graph_Mutant_K179C_MERS_nsp1}
\end{center}
\end{figure}

Figure~\ref{CL_Graph_Mutant_K164C_Cov2_nsp1}, \ref{CL_Graph_Mutant_K164C_Cov_nsp1},  \ref{CL_Graph_Mutant_K179C_MERS_nsp1} display CL signal graph for three mutants -
(a) K164C - original amino acid Lys (K) at serial location 164 mutated with amino acid Cys (C)) for CoV-2 nsp1 AA chain;
(b) K164C - original amino acid Lys (K) at serial location 164 mutated with amino acid Cys (C)) for CoV nsp1 AA chain;
(c) K179C - original amino acid Lys (K) at serial location 179 mutated with amino acid Cys (C) for MERS nsp1 AA chain. It can be observed that for each of the three mutant CL graph, an additional signal appears around cell locations 656 (4 x 164) corresponding to AA location 164 for CoV-2, CoV in Figure~\ref{CL_Graph_Mutant_K164C_Cov2_nsp1} and \ref{CL_Graph_Mutant_K164C_Cov_nsp1}. On the average, an AA is configured with 4 CA cells. The additional signal in Figure~\ref{CL_Graph_Mutant_K179C_MERS_nsp1} for nsp1 of MERS, appears beyond cell location 716 corresponding to AA location 179.

\subsubsection{Evaluation of Difference of CL Value Sum (DCLVS) between Wild and Mutant AA Chain Signal Graphs}
\label{DCLVS_wild_mutant}

\vspace{2mm}
CL Value Sum (CLVS) for a wild is computed considering values of mcl and Nmcl signal covering single/multiple cells of wild CL graph. Similarly, CLVS is computed out of a mutant CL graph. Next, we derive the CA model parameter DCLVS (Difference of CL Value Sum) defined as: 

~~~~~~~~~~~~~~~~~~DCLVS = (CLVS for mutant) - (CLVS for wild)

For evaluating DCLVS, we ignore the cells displaying CL value (-5).  For example, Figure~\ref{CL_Graph_Wild_AA_MERS_nsp1} and \ref{CL_Graph_Mutant_K179C_MERS_nsp1} display MERS wild and MERS mutant K179C, each with 7 cells having CL value (-5) around cell location 685; such cells are excluded for CLVS evaluation. Further, the mutant that shows -ive DCLVS value are excluded for mutational study in CAML model. A -ive DCLVS value indicates that some signal of wild get reduced/eliminated in the mutant. While a +ive DCLVS value of a mutant indicates appearance of some signal in the mutant CL graph that is absent in its wild counterpart. Appearance of such extra signal for a mutant has been found to be associated with the difference in the mutant function compared to that of wild. For example, in the CL signal graphs for three mutants in Figure~\ref{CL_Graph_Mutant_K164C_Cov2_nsp1}, \ref{CL_Graph_Mutant_K164C_Cov_nsp1},  \ref{CL_Graph_Mutant_K179C_MERS_nsp1}, a signal appears for mutants around the cell location 660 (CoV-2/CoV) and 735 (MERS). Such extra signal reported in a mutant and absent in wild represents deviation of mutant function in respect of its interaction with other biomolecules; presence of such a signal inhibits binding of the mutant with the target biomolecule, while no such inhibition in binding exists for the wild counterpart. 

The above considerations guide us to design the Machine Learning (ML) framework in the next section.

\section{Machine Learning (ML) - A Generic Framework for CAML Model}
\label{Generic_CAML}

The generic framework of Machine Learning enables us to map the physical domain features reported in published literature to the CAML model parameter DCLVS (GE value 50) derived for mutations inserted in a protein chain.                                                          
Table~\ref{AA_Composite_rule_Vector} shows 5 groups of AA (non-polar, polar uncharged, polar +ive, polar -ive, and aromatic ring). In order to                                                                                                                                                                                      identify the AA group that is most critical for structure and function of nsps, CAML model introduces 19 mutations for each AA of nsp chains and evaluate - (i) the count of non-0 DCLVS and (ii) count of +ive valued DCLVS.  The polar +ive group AAs displays highest count values across the nsps. Hence, CAML model assumes that polar +ive AA group react with biomolecules more than other four AA groups.  For subsequent discussions, a polar +ive AA is marked as Reactive AA (RAA).

The Machine Learning (ML) framework is designed from the exhaustive study and analysis of -                                                                                                    (a) in vitro/in vivo/in silico results surveyed in Section~\ref{survey_nsps}, in respect of  AAs - (i)  interacting with other biomolecules, or (ii) associated with enzymatic function; and                                                                                                                                                                         (b) the parameter DCLVS derived out of CAML model on inserting Cys mutations on the affected AAs identified in the survey (reported in Section~\ref{survey_nsps}).                                                                                                                                                                                                                                                            This framework enables us to ascertain the relationship of (b) and (a).

\subsection{ML Framework Design Methodology}
\label{ML_Fram_design_Method}

\vspace{2mm}
A few properties (P0 to P6) relevant for the ML framework design are highlighted prior to reporting the algorithmic steps of the design. These properties highlight the underlying principle followed in the ML framework design algorithm reported in next section.

\vspace{2mm}
\noindent
\textbf{P0 : CAML model for Mutational Study -} The wild CL graph of any protein is derived out the optimum AA chain for execution of its normal function. The wild AA chain has evolved out of natural evolution of the protein from its DNA/RNA. By inserting all possible 19 mutations for an amino acid in a protein chain, the CAML model evaluates importance of an amino acid relevant for execution of its function in respect of its interaction with other biomolecules or its enzymatic function. The mutant with (-ive) DCLVS value points to the fact that one or more signals in the mutant CL graph are absent that are present in wild CL graph; such mutants are not processed further for mutational study.  On the other hand, a (+ive) DCLVS value indicates that there are one or more signals present in mutant CL graph that are not present in wild CL graph. Consequently, such a mutant displaying +ive valued CL graph, is relevant for mutational study. Further, Cys-mutagenesis, as explained under Property P5, is identified on inserting 19 mutations for each AA.

\vspace{2mm}
\noindent
\textbf{P1 : The additional signal(s) for a mutant with +ive valued DCLVS confirms inhibition of binding of mutant with a biomolecule, while its wild with no such signal binds with the biomolecule -} As reported under Section~\ref{DCLVS_wild_mutant},  appearance of an extra signal for the mutant inhibits the interaction of the mutant with a biomolecule, whereas the CL signal graph of its  wild counterpart  having no such signal interacts with the biomolecule. Hence, the relevance of the AA displaying this characteristic gets established in CAML model in respect of binding of wild with a biomolecule, while the mutant fails to bind.

\vspace{2mm}
\noindent
\textbf{P2 : Importance of Polar +ive AA (Lys (K), Arg (R), and His (H))-} From analysis of AAs of nsps which interact with other biomolecules, or associated with its enzymatic function, it has been observed in Survey (Section~\ref{survey_nsps})  that polar +ive AAs are most active. Such an AA is marked as Reactive AA (RAA).

\vspace{2mm}
\noindent
\textbf{P3 :} Two adjacent RAA in nsp chain is marked as RAA pair. A sequence of 5 adjacent AAs covering at least two RAAs is marked as RAA region.

\vspace{2mm}
\noindent
\textbf{P4 : Acceptable range of count of non-0 DCLVS GE 50 -} From analysis of in vitro/in vivo/in silico experimental results and DCLVS values of mutants, the lower acceptable limit of DCLVS is set as greater than or equal to value 50.

\vspace{2mm}
\noindent
\textbf{P5 : Cys-mutagenesis for CAML model - } In order to identify specific AA location important for protein function, CAML model employs Cys-mutagenesis based on the following consideration. It has been observed that Cys mutation for different AA locations generates highest count of non-0 +ive valued DCLVS with value GE (greater than or equal to) 50 across the nsp chains of CoV-2, CoV, MERS.

\vspace{2mm}
\noindent
\textbf{P6 : Select a sequence of 5 AA as a potential candidate region for binding with a biomolecule -} CAML model assumes that 5 AAs covering a RAA pair (or RAA region) is critical for interaction of nsp with a biomolecule (protein or RNA chain). The 5 AAs are assumed to make bonds for binding with a biomolecule. A sequence of 5 AA covering a RAA pair (or RAA region) is identified and the sum of DCLVS values for the sequence is evaluated.
                                                                                                                                              
The 5 AA sequence with DCLVS sum GE 300 (average value GE 50) and no individual AA having value GE 300 is accepted as a potential candidate region for binding with a biomolecule.

\subsection{Design of ML Framework to derive CAML model results for nsps}
\label{ML_Framework_Design}

\vspace{2mm}
The algorithmic steps to design the Machine Learning (ML) framework of CAML model is reported below for an input protein chain of Group1 nsps - (i) on introducing Cys mutation for each AA of a protein chain (Ala mutation if the AA in protein chain is Cys), and evaluating DCLVS (Difference between wild and mutant CLVS).

\subsubsection{The Machine Learning (ML) Algorithm}

\noindent
\rule[5pt]{1.00\textwidth}{1.00pt}
\textbf{Input1 :} Amino acid (AA) chain of CoV-2, CoV, and MERS of Group1 nsps.\\
\textbf{Input2 : (from Section II Survey)} Detailed analysis of published literature on experimental in vivo/in vitro/in silico results in respect of - (i) binding of nsp with other biomolecule; and (ii) AAs associated with catalytic function of nsp.  \\
\rule[5pt]{1.00\textwidth}{1.00pt}\\
\textbf{Step 1 :} Generate CL signal graph for wild and mutants on introducing Cys mutation for each AA of the chain; for Cys AA in the sequence, introduce Ala mutation. Note the DCLVS value for each mutant. 

\vspace{2mm}
\noindent
\textbf{Step 2 :} For enzymatic nsps other than nsp1 and nsp6, go to Step 7.

\vspace{2mm}
\noindent
\textbf{Step 3(a) :} Identify a pair of adjacent AAs marked as RAA (Reactive AA) pair for CoV-2 and CoV AA sequence.

\vspace{2mm}
\noindent
\textbf{Step 3(b) :} For MERS, identify a sequence of 5 adjacent AAs in NTD N-terminal Domain) \& CTD (C terminal Domain) of nsp1 covering at least two polar positive AAs.

\vspace{2mm}
\noindent
\textbf{Step 4 :} Note the sequence of 5 AA around RAA pair for CoV-2 and CoV nsp1 with maximum DCLVS sum for the 5 AAs.

\vspace{2mm}
\noindent
\textbf{Step 5 :} Mark the 5 AA sequence as the binding regions for CoV-2, CoV, and MERS nsp1 with a biomolecule. 

\vspace{2mm}
\noindent
\textbf{Step 6 :} Compare DCLVS sum for the binding regions identified in Step 5 for CoV-2 and CoV nsp1. Go to Stop.

\vspace{2mm}
\noindent
\textbf{Step 7 :} Identify the catalytic domain AA and other functionally important AAs associated with enzymatic function. Introduce Cys mutation for each AA and Ala mutation for a Cys AA in the AA chain.

\vspace{2mm}
\noindent
\textbf{Step 8 :} Note the DCLVS values for each AA of the domain identified in Step 7 for CoV-2, CoV, and MERS to compare DCLVS sum for the functionally important domain(s) of CoV-2 and CoV. 

\vspace{2mm}
\noindent
\textbf{Step 9 :} If majority of the important AAs display (-ive) DCLVS value, compare the CLVS for CoV-2 and CoV wild CL graph to compare their contribution towards viral load and transmissibility.

\vspace{2mm}
\noindent
\textbf{Step 10 :} Stop. \\
\rule[5pt]{1.00\textwidth}{1.00pt}

The results derived for CoV-2, CoV, and MERS nsps on executing algorithmic steps are reported in next two sections. The leader protein nsp1 stands apart from other Group1 nsps in respect of interaction with other biomolecules for viral pathogenesis; next section reports the results for nsp1. Results for other group1 nsps are presented in Section~\ref{result_other_nsps}.

\section{Results Derived out of CAML model for nsp1 on executing ML Algorithm}
\label{result_nsp1}

Section~\ref{Task_nsp1} presents a brief survey on nsp1 protein of CoV-2, CoV, and MERS.  The tasks associated with execution of function of nsp1 are reviewed in the survey. The CAML model results derived for nsp1 are reported for three tasks (Ta1, Ta2, Ta3) associated with nsp1 function.

\subsection{Task Ta1}
\label{nsp1_Task_Ta1}

\begin{figure}[h!]
\begin{center}
\includegraphics[width=13.3cm, height=4.6cm]{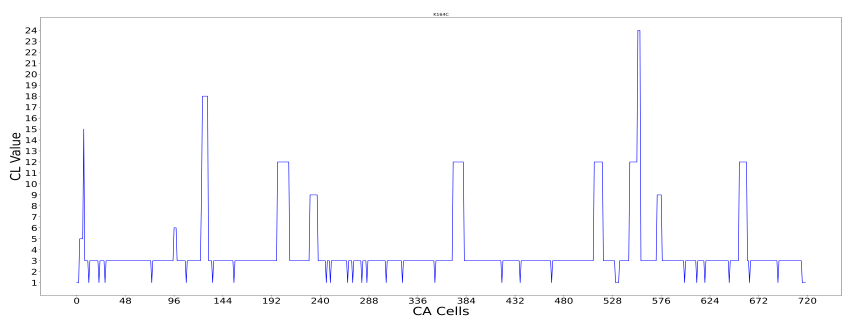}
\caption{CL signal graph for CoV-2 mutant K164C with high valued DCLVS value 96 at CTD end around the cell location 660 corresponding to amino acid K164}
\label{CL_Graph_nsp1_Cov2_Mutant_K164C}
\end{center}
\end{figure}

\begin{figure}[h!]
\begin{center}
\includegraphics[width=13.3cm, height=4.6cm]{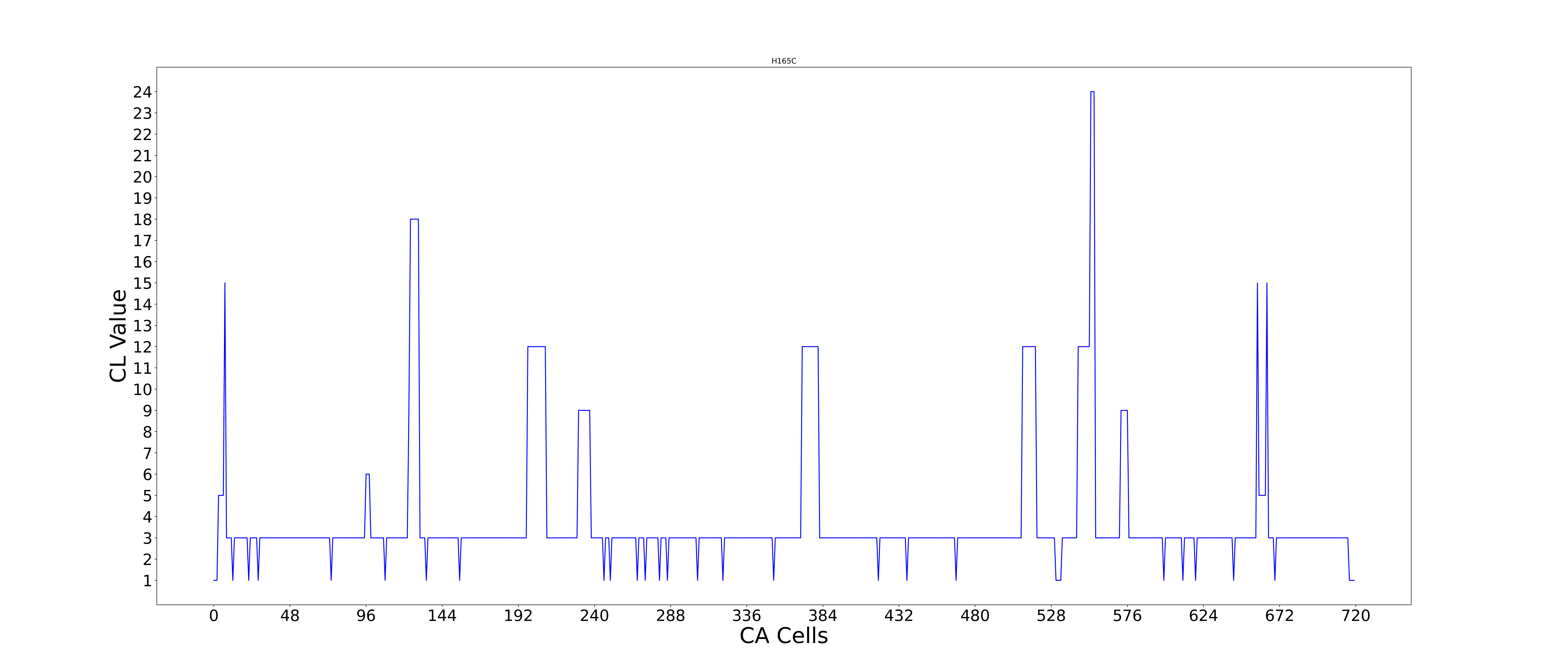}
\caption{CL signal graph for CoV-2 mutant H165C with  DCLVS value 55 at CTD end around the cell location 660 corresponding to amino acid K164}
\label{CL_Graph_nsp1_Cov2_Mutant_H165C}
\end{center}
\end{figure}

\vspace{2mm}
Results derived out of CAML model in respect of the RAA pair (K164, H165) and the associated 5 AA sequence are reported in Table~\ref{5_AA_DCLVS}.  Figure~\ref{CL_Graph_Wild_AA_Cov2_nsp1}  and \ref{CL_Graph_Wild_AA_Cov_nsp1} show the CL signal graph for wild CoV-2 and CoV, while Figure~\ref{CL_Graph_nsp1_Cov2_Mutant_K164C} and ~\ref{CL_Graph_nsp1_Cov2_Mutant_H165C} reports CL signal graphs for two mutants K164C and H165C of CoV-2. Both show a high valued signal above value 50 at C terminal end of nsp1 around cell location 660 corresponding to AA 164 and 165. Such a signal, as explained in Section~\ref{DCLVS_wild_mutant}, inhibits binding of nsp1 with a biomolecule. Similar results are also derived for CoV mutants K164C and H165C. The wild CL signal graphs of CoV-2 and CoV (Figure~\ref{CL_Graph_Wild_AA_Cov2_nsp1}  and \ref{CL_Graph_Wild_AA_Cov_nsp1}) do not show such a signal at CTD end; hence wild nsp1 binds on the concerned biomolecule.

As per Property P6, the 5 AA sequence covering RAA (K164, H165), derived in Step 4, is a potential candidate for binding with the biomolecule - host ribosome 40S subunit reported in published literature.

\vspace*{-3mm}

\begin{table}[h!]
\centering
\resizebox{0.8\textwidth}{!}{  
\begin{tabular}{|c|ccccc|c|c|}
\hline 
$1^{st}$& \multicolumn{5}{c|}{DCLVS value for 5 AA} & Sum of  & Average \\ 
option & W161 & N162 & T163 & K164 & H165 & DCLVS & value \\ 
\hline 
CoV & 0 & 75 & 195 & 75 & 120 & 465 & $465/5 = 93$ \\ 
\hline 
CoV-2 & 84 & 200 & 270 & 96 & 55 & 705 & $705/5 = 141$ \\ 
\hline 
$2^{nd}$& \multicolumn{5}{c|}{DCLVS value for 5 AA} &  &  \\ 
option & K164 & H165 & S166 & S167 & G168 &  &  \\ 
\hline 
CoV & 75 & 120 & 0 & 0 & 147 & 342 & $342/5 = 68.40$ \\ 
\hline 
CoV-2 & 96 & 55 & 0 & 0 & 252 & 403 & $404/5 = 80.60$ \\ 
\hline 
\end{tabular} }
\caption{\small{DCLVS values of two 5 AA sequences showing DCLVS sum for CoV-2 and CoV}}
\label{5_AA_DCLVS} 
\end{table}

As per Step 4, there are two options for 5 AA sequence as (R161 to H165) and (K165 to G168). The data for both the options are reported in Table~\ref{5_AA_DCLVS}. The first option is selected, as per Step 4, since it generates higher value sum. It is accepted as the binding region of nsp1 with host ribosome.

Table~\ref{5_AA_DCLVS} results confirm higher degree of binding affinity for CoV-2 compared to that of CoV for binding with host ribosome. Published experimental results reported in \cite{shen2021lysine, schubert2020sars, nakagawa2021mechanisms, simeoni2021nsp1, terada2017mers, benedetti2020emerging, clark2021structure, kumar2020sars, tidu2021viral, vankadari2020structure} match with the result derived out of CAML model.

\subsection{Task Ta2}
\label{nsp1_Task_Ta2}

\begin{figure}[h!]
\begin{center}
\includegraphics[width=13.3cm, height=4.5cm]{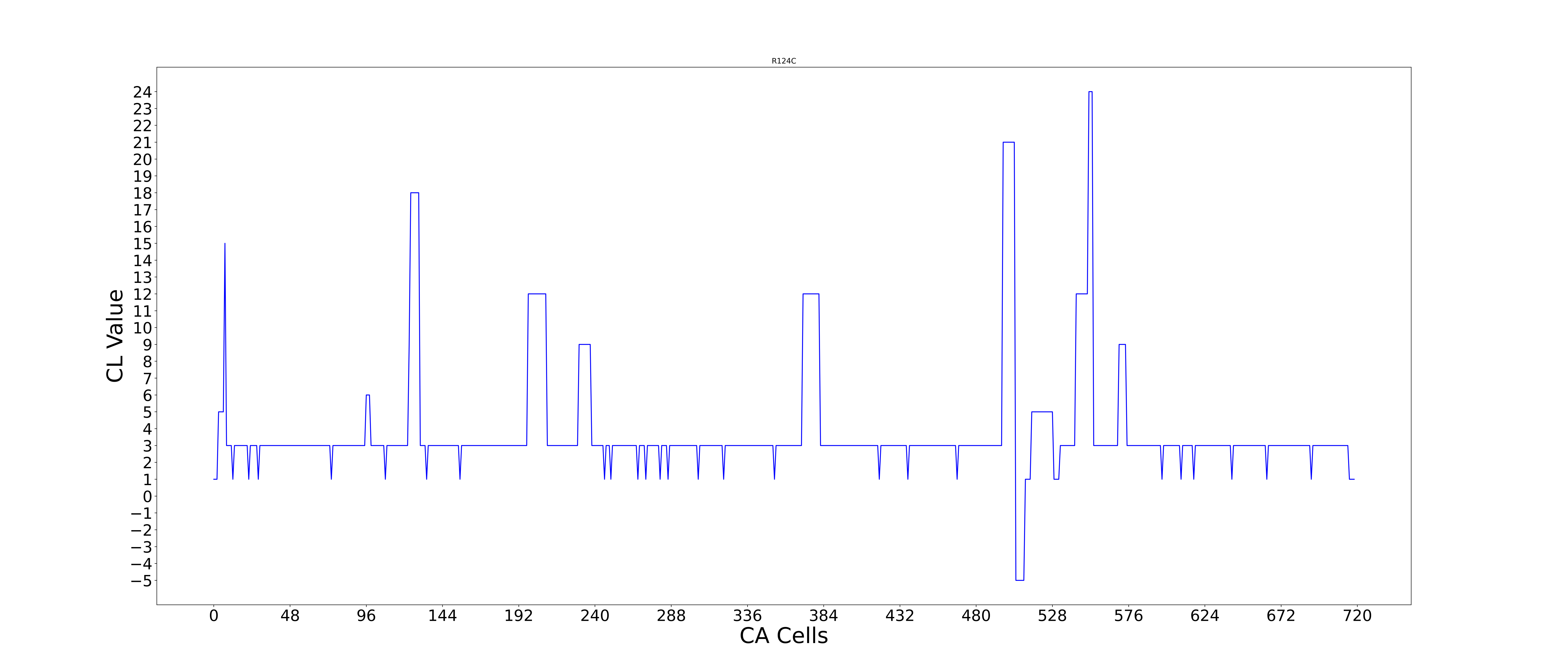}
\caption{CL signal graph for the mutant R124C of CoV-2 displaying a high valued signal around cell location 500 corresponding to AA location 124 that is absent in the CoV-2 wild CL graph reported in Fig 5(a)}
\label{CL_Graph_nsp1_Cov2_Mutant_R124C}
\end{center}
\end{figure}

\begin{figure}[h!]
\begin{center}
\includegraphics[width=13.3cm, height=4.7cm]{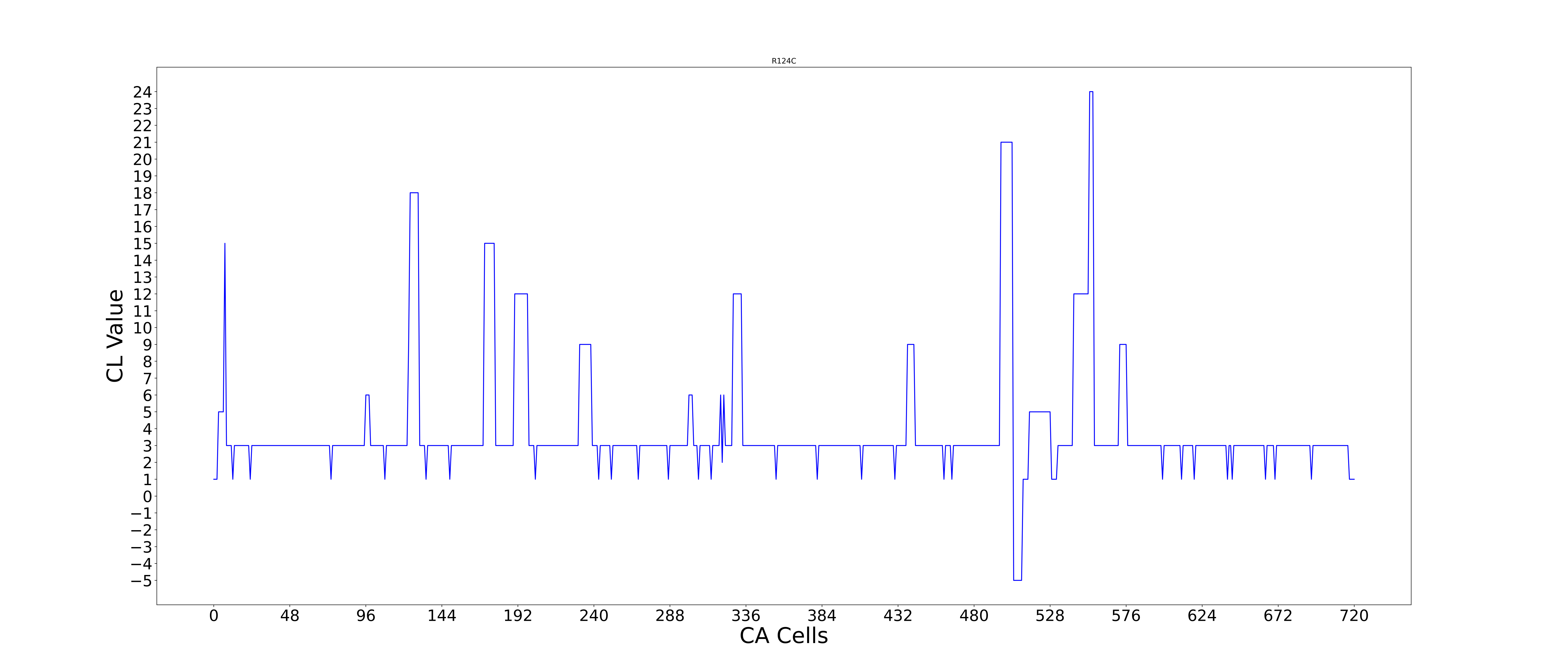}
\caption{CL signal graph for the mutant R124C of CoV displaying a high valued signal around cell location 500 corresponding to AA location 124 that is absent in the CoV wild CL graph reported in Fig 5(b)}
\label{CL_Graph_nsp1_Cov_Mutant_R124C}
\end{center}
\end{figure}

\vspace{2mm}
Results derived out of CAML model in respect of RAA pair (R124 and K125) and the associated 5 AA sequence are reported in Table~\ref{5_AA_DCLVS_Ta2}. Figure~\ref{CL_Graph_nsp1_Cov2_Mutant_R124C} and ~\ref{CL_Graph_nsp1_Cov_Mutant_R124C} reports CL signal graphs for mutant R124C for CoV-2 and CoV. The 5 AA sequence covering the RAA pair is noted with average DCLVS value as 63 for both CoV-2 and CoV. Hence the 5 AA sequence V121 to K125 is a valid nsp1 candidate 5 AA sequence for binding with a biomolecule (viral 5'UTR Stem Loop SL1).  As per the results of Table~\ref{5_AA_DCLVS_Ta2}, the efficiency of execution of  task Ta2 are identical for CoV-2 and CoV.

\begin{table}[h]
\centering
\begin{tabular}{|c|ccccc|c|c|}
\hline 
 & \multicolumn{5}{c|}{DCLVS value for 5 AA} & Sum of  & Average \\ 
  & V121 & L122 & L123 & R124 & K125 & DCLVS & value \\ 
\hline 
CoV-2 & (-78) & 65 & 189 & 60 & (-108) & 314 & 63 \\ 
\hline 
CoV & (-102) & 65 & 189 & 60 & (-108) & 314 & 63 \\ 
\hline 
\end{tabular} 
\caption{DCLVS values of 5 AA sequence showing DCLVS values showing DCLVS sum 63 for CoV-2 and CoV}
\label{5_AA_DCLVS_Ta2}
\end{table}

\subsection{Task Ta3}
\label{nsp1_Task_Ta3}

\vspace{2mm}
As per the Observation noted in Section~\ref{Task_nsp1} out of nsp sequence analysis and published experimental results, a pair of adjacent polar positive AAs (Lys K, Arg R, His H) of nsp sequence play critical role for interaction with biomolecules. For example, the RAA pair (K164, H165) and (R124, K125) are associated for interaction with host ribosome and viral 5'UTR, as noted for tasks Ta1 and Ta2. Cov-2 nsp1 displays another RAA pair (R119, K120). Based on the results in \cite{terada2017mers, lokugamage2015middle} in respect of possible cleavage function of MERS, it has been assumed that 5 AA sequence with RK pair (119, 120) of CoV-2 interact with proteins of host defense immune system to down regulate host immune response.

\begin{figure}[h!]
\begin{center}
\includegraphics[width=13.3cm, height=4.7cm]{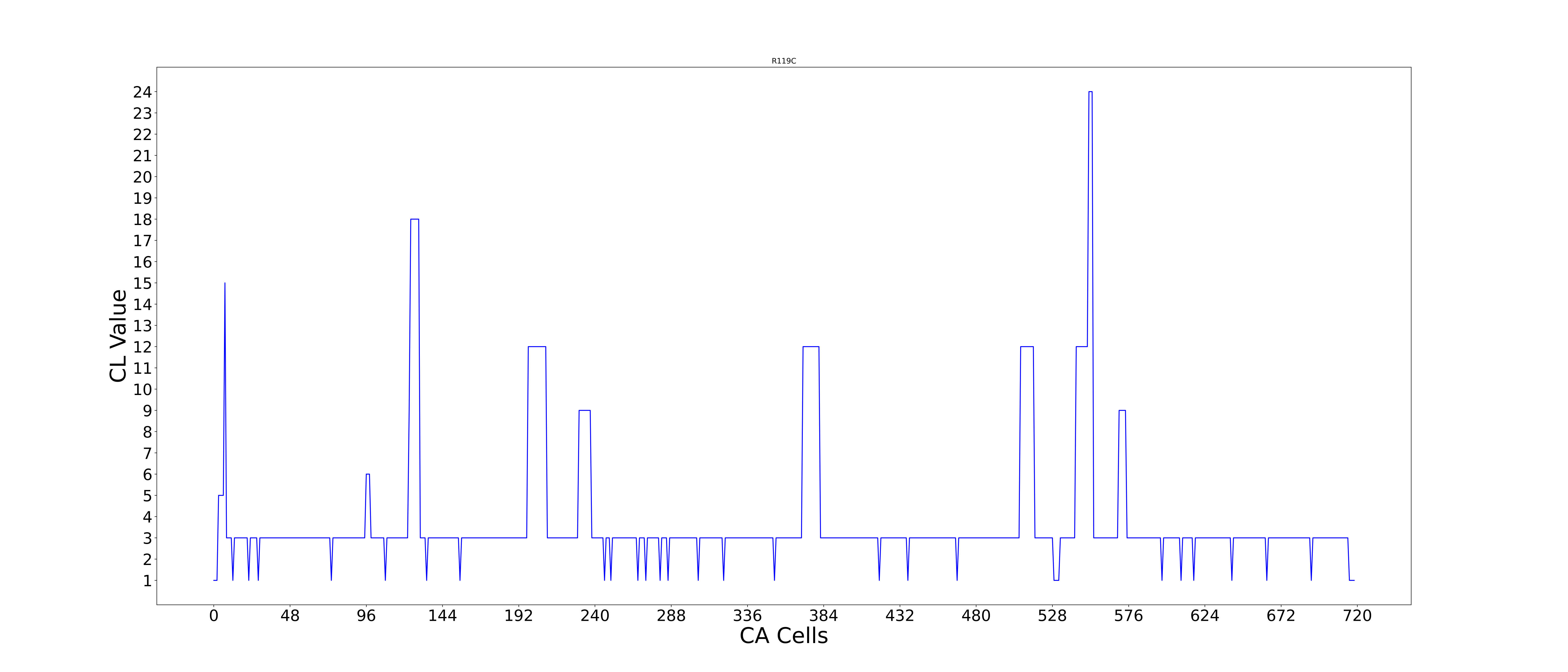}
\caption{CL signal graph for mutant R119C - it is similar to that of wild (Fig 5(a))}
\label{CL_Graph_nsp1_Cov2_Mutant_R119C}
\end{center}
\end{figure}

\begin{figure}[h!]
\begin{center}
\includegraphics[width=13.3cm, height=4.7cm]{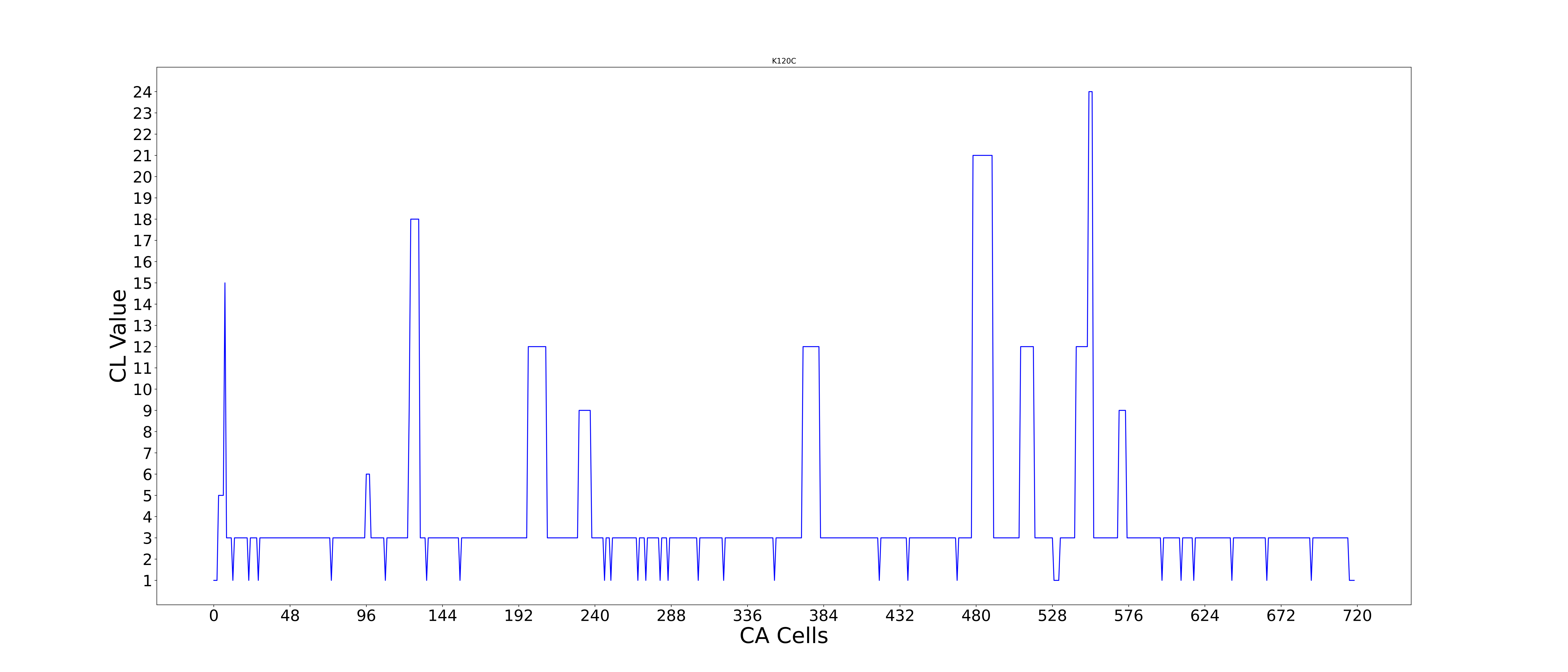}
\caption{CL signal graph for CoV-2 mutant K120C showing a large signal of value around cell location 475 corresponding to AA location 120}
\label{CL_Graph_nsp1_Cov2_Mutant_K120C}
\end{center}
\end{figure}

Results derived out of CAML model in respect of RAA pair (R119, K120) of CoV-2 and the associated 5 AA sequence are reported in Table~\ref{5_AA_DCLVS_Ta3}. Figure~\ref{CL_Graph_nsp1_Cov2_Mutant_R119C} and \ref{CL_Graph_nsp1_Cov2_Mutant_K120C} show the CL signal graphs for CoV-2 mutants R119C and K120C. While CL graphs of mutant R119C is identical to that of CoV-2 wild (Figure~\ref{CL_Graph_Wild_AA_Cov2_nsp1}), the mutant K120C displays a high valued signal. The 5 AA sequence (119 to 123) covering CoV-2 RAA (R119, K120) displays DCLVS sum as 527 with average value 105. Hence CoV-2 AA sequence (119 to 123) is a potential candidate for binding with a biomolecule. However, this observation is not true for CoV that shows (R119, N120) with 5 AA covering R119 showing total value as 254 and average value 51.

\begin{table}[h]
\centering
\begin{tabular}{|c|ccccc|c|c|}
\hline 
 & \multicolumn{5}{c|}{DCLVS value for 5 AA} & Sum of  & Average \\ 
 & R119 & K120 & V121 & L122 & L123 & DCLVS & value \\ 
\hline 
CoV-2 & 0 & 273 & (-78) & 65 & 189 & 527 & $527/5 = 105$ \\ 
\hline 
 & \multicolumn{5}{c|}{DCLVS value for 5 AA} &  &  \\ 
 & R119 & N120 & V121 & L122 & L123 &  &  \\ 
\hline 
CoV & 0 & (-300) & (-102) & 65 & 189 & 254 & $254/5 = 51$ \\ 
\hline 
\end{tabular} 
\caption{DCLVS values for 5 AA sequence showing DCLVS sum for CoV-2 and CoV}
\label{5_AA_DCLVS_Ta3}
\end{table}

CAML model assumes that the 5 AA sequence (119 to 123) of nsp1 around RAA pair (R119, N120) is associated for execution of task Ta3 to inhibit/restrict host immune defense. Table~\ref{5_AA_DCLVS_Ta3} results confirm higher efficiency for execution of Ta3 for CoV-2 compared to that of CoV.

\subsection{Summary of results derived out CAML model}

\vspace{2mm}
In summary, CoV-2 executes the primary task Ta1 (host translation shut off) with higher efficiency compared to CoV, since the average DCLVS values are respectively 141 and 93 (50 \% higher) for CoV-2 and CoV respectively. The higher efficiency of host translational shut off will enable the SARS CoV-2 virus to replicate and transmit with higher efficiency than SARS CoV. Further, this higher transmissibility of CoV-2 gets strengthened due to execution of task Ta3 in respect of better evasion of host immune defense than that of CoV; the respective values are 105 and 51 (100 \% higher) for CoV-2 than CoV.

\noindent
We next report the results for MERS nsp1 to investigate its differences with CoV nsp1.

\subsection{Results derived out of CAML model for MERS nsp1}
\label{MERS_nsp1_Result}

\vspace{2mm}
For design of CAML model, a crucial observation is reported in Section~\ref{Task_nsp1} under the head `\emph{Observation from Analysis of nsp sequences and Published Experimental Results}'. The observation refers to the key role played by polar positive amino acids in SARS and MERS nsps. In order to execute Step 3(b) of ML algorithm, we identify the polar positive amino acids in MERS nsp1 sequence with maximum distance of 5 between two AAs as: (R13, R17), (K24, H25), (K65, K66), (R146, K147), (K179, K181), (K188, K189). We align 193 AA sequence of MERS with 180 AA of CoV-2/CoV to identify AAs in MERS nsp1 equivalent to AA pair (K168, H169 - binding with host ribosome, and R13 - for recognition of RNA 5'UTR Stem Loop SL1. In the process we identify -  (i) 5 AA sequence L177 to K181 covering two polar +ive AAs (K179, K181) in CTD, and (ii) 5 AA sequence (R13 to R17) covering two polar positive AAs  R13, R17 in  NTD of MERS nsp1. These two 5 AA sequences are associated with tasks Ta1 and Ta2 of MERS nsp1. For the sake of comparison with CoV, the corresponding average values are copied from Table~\ref{5_AA_DCLVS} (Section~\ref{nsp1_Task_Ta1}), Table~\ref{5_AA_DCLVS_Ta2} (Section~\ref{nsp1_Task_Ta2}), and Table~\ref{5_AA_DCLVS_Ta3} (Section~\ref{nsp1_Task_Ta3}).

\subsubsection{Task (Ta1) derived out of CAML model for MERS nsp1}

\begin{figure}[h!]
\begin{center}
\includegraphics[width=13.3cm, height=4.7cm]{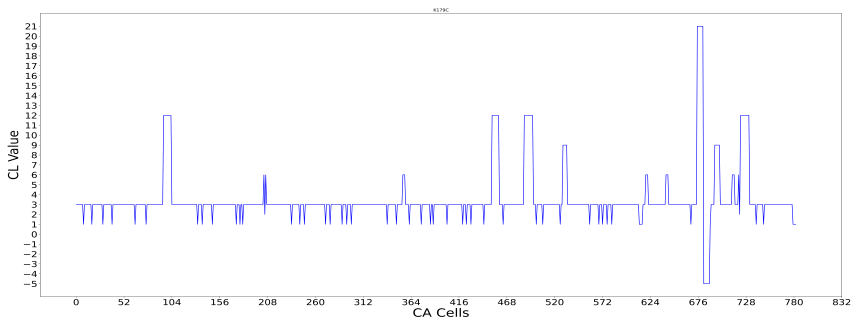}
\caption{CL signal graph for mutant K179C showing high valued signal around cell location 725 corresponding to K179 for task Ta1}
\label{CL_Graph_nsp1_MERS_Mutant_K179C}
\end{center}
\end{figure}

\vspace{2mm}
The 5 AA sequence (L177 to K181) selected around AA in location (K179, K181) is noted in Table~\ref{5_AA_DCLVS_MERS_Ta1} with DCLVS value for Cys mutant of each AA. The sum of DCLVS values and average value are 603 and 121. Hence the region of 5 AA sequence (177 to 181) is identified as a potential candidate for binding on a biomolecule. The CL signal graph for mutant K179C is reported in Figure~\ref{CL_Graph_nsp1_MERS_Mutant_K179C} with high valued signal around cell location 725 corresponding to AA location 179. Hence, 5 AA sequence (177 to 181) is identified as the region of MERS nsp1 binding with host ribosome to shut down host mRNA translation.

\begin{table}[h]
\centering
\resizebox{\textwidth}{!}{  
\begin{tabular}{|ccccc|c|c|c|}
\hline 
 \multicolumn{5}{|c|}{DCLVS value for 5 AA} & Sum of  & Average  & Average value \\ 
 L177 & N178 & K179 & G180 & K181 & DCLVS & value (MERS) & (CoV from Table~\ref{5_AA_DCLVS}) \\ 
\hline 
168  & 180  & 120 & 135 & 0 & 603 &  121 (23 \% higher than CoV) & 93 \\ 
\hline 
\end{tabular} }
\caption{DCLVS values for AAs associated with execution of task Ta1 for MERS}
\label{5_AA_DCLVS_MERS_Ta1}
\end{table}

\subsubsection{Task (Ta2) derived out of CAML model for MERS nsp1}

\vspace{2mm}

\begin{figure}[h!]
\begin{center}
\includegraphics[width=13.3cm, height=4.7cm]{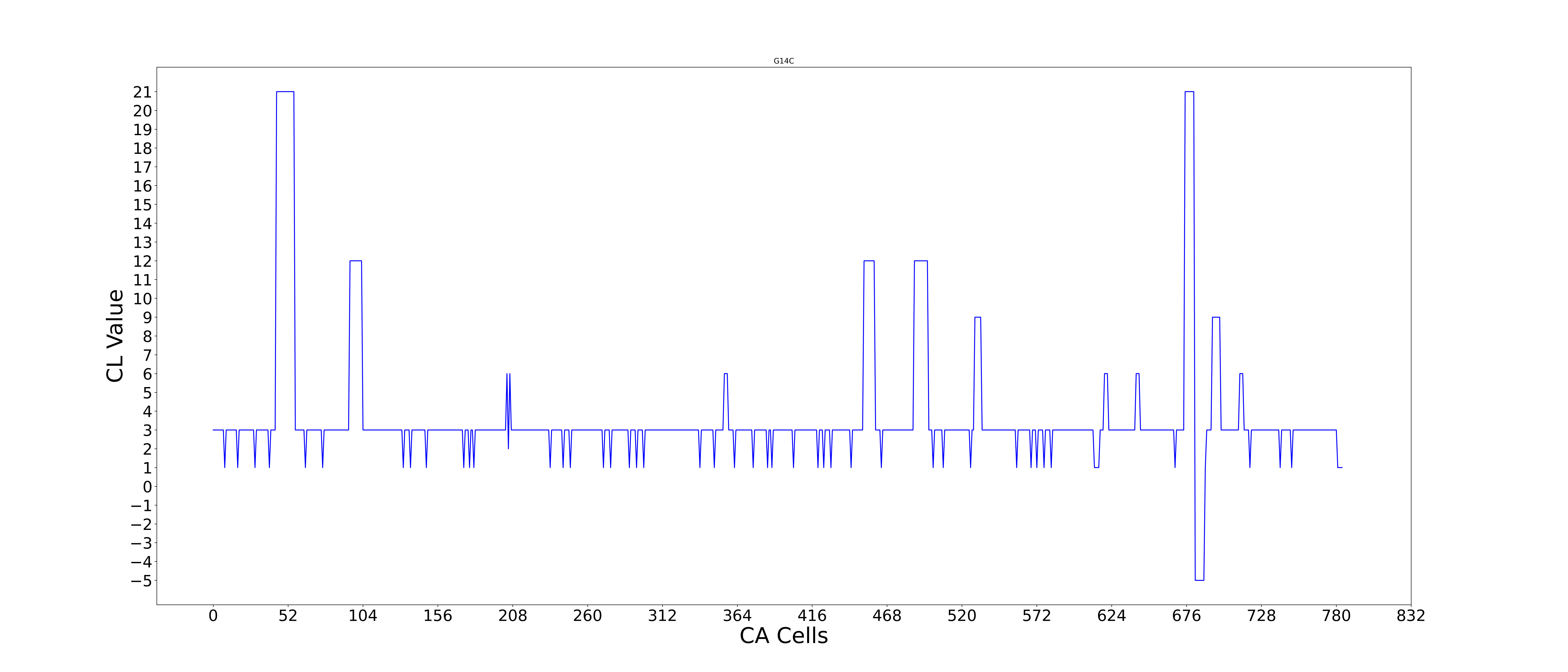}
\caption{MERS CL signal graph for mutant G14C showing high valued signal around cell location 55 corresponding to G14 for task Ta2}
\label{CL_Graph_nsp1_MERS_Mutant_G14C}
\end{center}
\end{figure}

The 5 AA sequence (R13 to R17)  location  covering polar +ive AA (R13, R17) is noted in Table~\ref{5_AA_DCLVS_MERS_Ta2} with DCLVS value for Cys mutant of each AA. The sum of DCLVS values and average value are 403 and 81.  Hence the region of 5 AA (13 to 17) sequence is identified as the potential candidate for binding on 5'UTR of MERS viral RNA. The MERS CL signal graph for mutant G14C is reported in Figure~\ref{CL_Graph_nsp1_MERS_Mutant_G14C} with high valued signal around cell location 55 corresponding to AA location 14.

\begin{table}[h]
\centering
\resizebox{\textwidth}{!}{  
\begin{tabular}{|ccccc|c|c|c|}
\hline 
 \multicolumn{5}{|c|}{DCLVS value for 5 AA} & Sum of  & Average  & Average value \\ 
 R13 & G14 & T15 & Y16 & R17 & DCLVS & value (MERS) & (CoV from Table~\ref{5_AA_DCLVS_Ta2}) \\ 
\hline 
 0 & 273  & 65 & 65 & 0 & 403 &  81 (22 \% higher than CoV) & 63 \\ 
\hline 
\end{tabular} }
\caption{DCLVS values for AAs associated with execution of task Ta2 for MERS}
\label{5_AA_DCLVS_MERS_Ta2}
\end{table}

\subsubsection{Task (Ta3) derived out of CAML model for RAA pair (R146, K147)}

\begin{figure}[h!]
\begin{center}
\includegraphics[width=13.3cm, height=4.7cm]{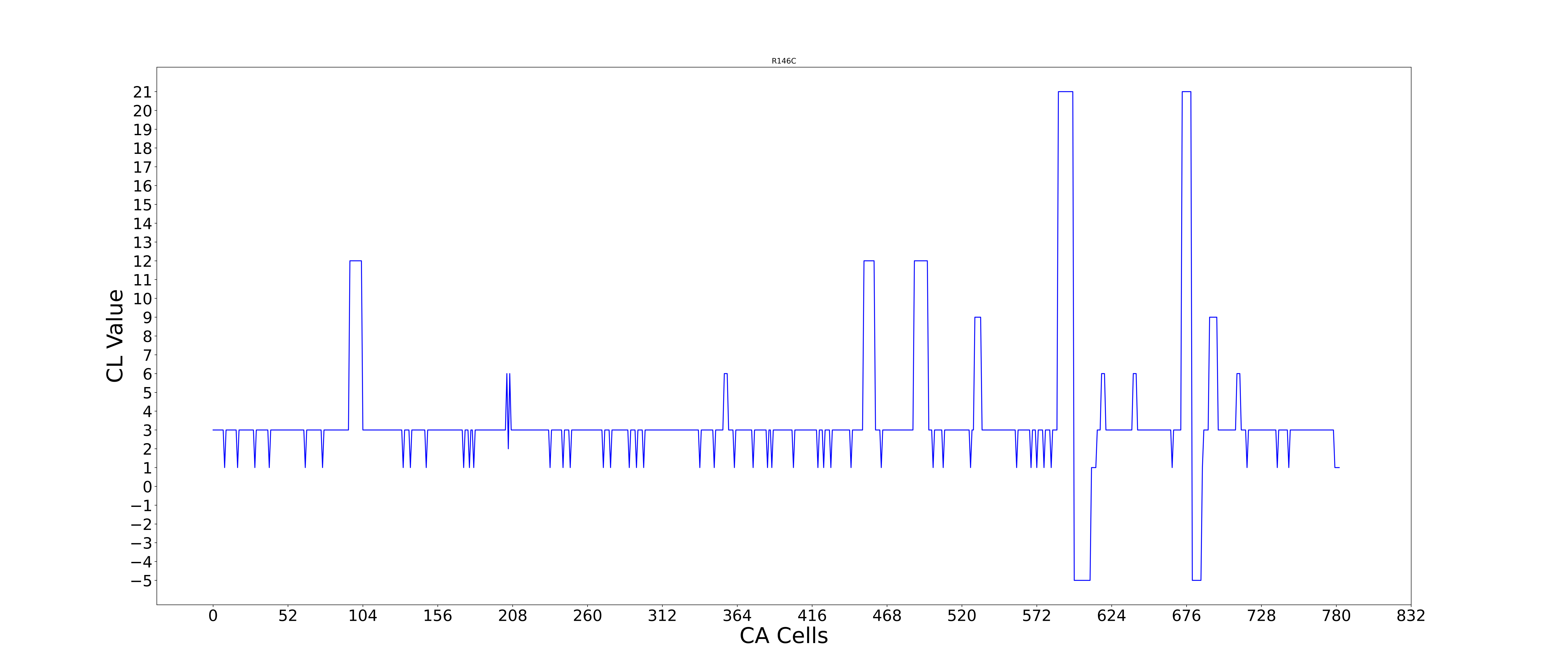}
\caption{CL signal graph for mutant R146C showing high valued signal around cell location 585 corresponding to R146 for task Ta3}
\label{CL_Graph_nsp1_MERS_Mutant_R146C}
\end{center}
\end{figure}

\begin{figure}[h!]
\begin{center}
\includegraphics[width=13.3cm, height=4.7cm]{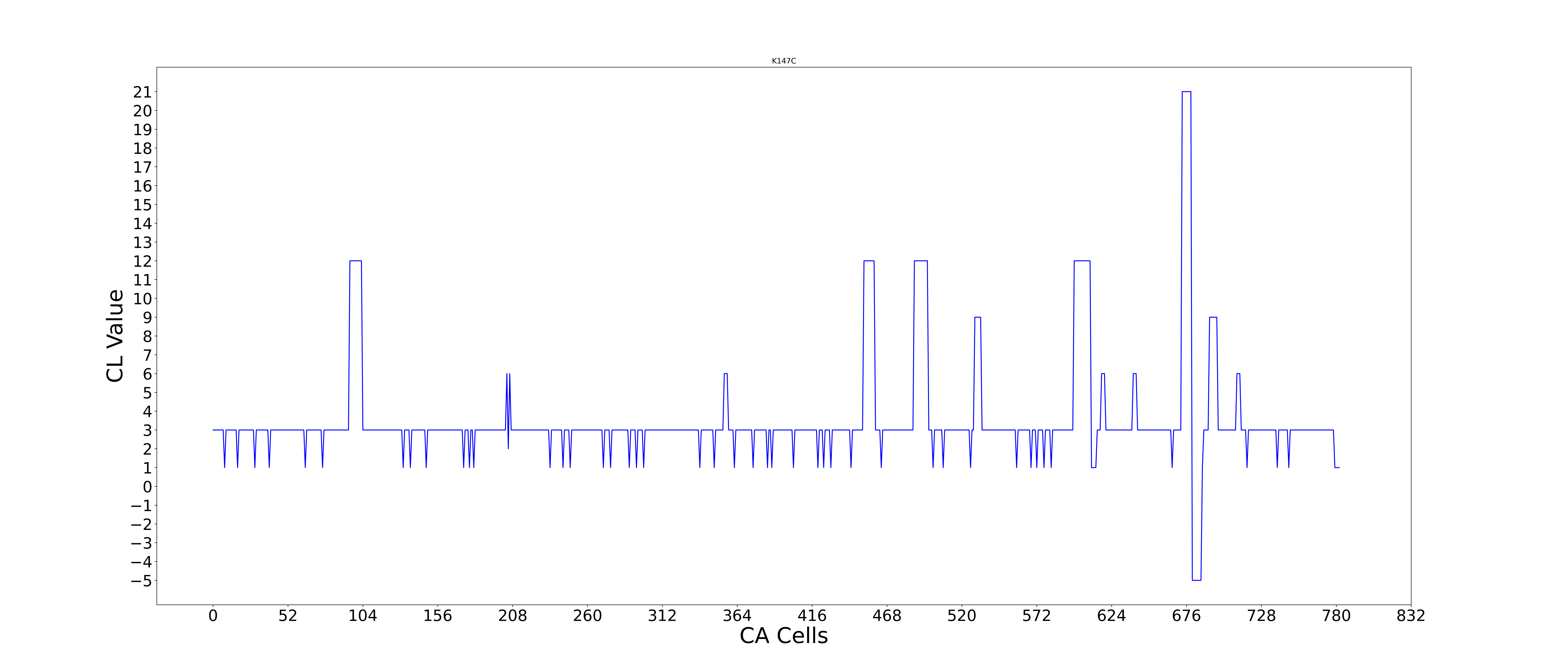}
\caption{CL signal graph for mutant K147C showing high valued signal around cell location 590 corresponding to K147 for task Ta3}
\label{CL_Graph_nsp1_MERS_Mutant_K147C}
\end{center}
\end{figure}

\vspace{2mm}
It has been reported \cite{terada2017mers, lokugamage2015middle} that in addition to host translation shut off, MERS nsp1 displays host RNA cleavage function with AA pair (R146, K147). The 5 AA sequence (R146 to R150) selected around RAA pair (R146, K147) is noted in Table~\ref{5_AA_DCLVS_MERS_Ta3} with DCLVS value for Cys mutant of each AA. The sum of DCLVS values and average value are 531 and 106.  Hence the region of 5 AA sequence (146 to 150) is identified as a potential candidate for binding on a biomolecule. The CL signal graph for mutant R146C and K147C are reported in Figure~\ref{CL_Graph_nsp1_MERS_Mutant_R146C} and \ref{CL_Graph_nsp1_MERS_Mutant_K147C} with high valued signal around cell location 588 corresponding to AA location 146 and 147.  The 5 AA sequence (146 to 150) is associated with RNA cleavage activity.  As a result, the proteins necessary for activation of host immune response are not synthesized leading to compromised host immune response due to MERS nsp1.

\begin{table}[h]
\centering
\resizebox{\textwidth}{!}{  
\begin{tabular}{|ccccc|c|c|c|}
\hline 
 \multicolumn{5}{|c|}{DCLVS value for 5 AA} & Sum of  & Average  & Average value \\ 
 R146 & K147 & Y148 & G149 & R150 & DCLVS & value (MERS) & (CoV from Table~\ref{5_AA_DCLVS_Ta3}) \\ 
\hline 
 231 & 144 & 156 & 0 & 0 & 531 & 106 (100 \% higher than CoV) & 51 \\ 
\hline 
\end{tabular} }
\caption{DCLVS values for AAs associated with execution of task Ta3 for MERS}
\label{5_AA_DCLVS_MERS_Ta3}
\end{table}

\subsubsection{Summary of results derived out of CAML model for comparison of Transmissibility and Virulence of SARS Covid (2003) and MERS covid (2012)}

\vspace{2mm}
In view of limited data of affected patients (in terms of a few thousands) it is not possible to compare transmissibility of CoV and MERS. However, due to significantly higher (100 \% higher) value of task Ta3 (in respect of suppression of host immune response), virulence for MERS is likely to be higher than CoV. This results of CAML model gets validated from the historical records reproduced below:                                                                                                                      

The number of (affected countries, infected persons, death count) for 2003 SARS covid (CoV) and 2012 MERS covid are respectively (29, 8000, 774) and (27, 2500, 876) with ratio (of death count and affected persons) as 9.6 \% for CoV and 35.0 \% for MERS. Higher percent of death count can be attributed to higher host immune suppression triggering higher virulence.

\section{Results derived out of CAML Model for other nsps}
\label{result_other_nsps}

Out of the remaining nsps, all the nsps excepting nsp6 have enzymatic function.

\subsection{CAML results on nsp3 - PLpro encoded within nsp3}
\label{CAML_nsp3_result}

\vspace{2mm}
As per the survey presented in Section~\ref{Task_nsp3}, Plpro encoded within nsp3, plays a prominent role for cleavage of viral RNA - a fundamental step in virus life cycle. Important conserved amino acids of PLpro are -\\                                                                                                                                                                                                                                                 (i) Catalytic triads (C111, H272, and D286);\\                                                                                                                                                                                                   (ii) other catalytically important AAs : W93, W106, D108, N109 (L106 in MERS); and \\                                                         (iii) hydrophobic cleft (P247, P248, Y264) common for CoV-2, CoV, and MERS.

We employ CAML model to identify DCLVS between CoV-2 and CoV PLpro, with Cys mutations introduced for each of the important amino acids. The DCLVS values for introduction of Cys mutations (or Ala mutation if AA is Cys) are reported in Table~\ref{DCLVS_nsp3}.

\begin{table}[h!]
\centering
\resizebox{\textwidth}{!}{  
\begin{tabular}{|c|c|c|c|c|c|c|c|c|c|c|c|c|}
\hline 
 & \multicolumn{3}{|c|}{Catalytic triads} & \multicolumn{4}{|c|}{Other catalytically important AAs} &  & \multicolumn{3}{|c|}{Hydrophobic cleft AAs} &  \\ 
\hline 
Virus & C111A & H272C & D286C & W93C & W106C & D108C & N109C & Sum & P247C & P248C & Y264C & Sum \\ 
\hline 
CoV-2 & (-180) & (-54) & (-18) & 117 & 105 & 78 & (-339) & 300 & 147 & 126 & 291 & 564 \\ 
\hline 
CoV & (-198) & 63 & 102 & 135 & 0 & (-168) & (-294) & 135 & 0 & (-42) & 171 & 171 \\ 
\hline 
MERS & (-180) & (-96) & (-72) & 105 & ** & 12 & 27 & 105 & (-204) & (-204) & 0 & 0 \\ 
\hline 
\end{tabular} }
\begin{flushleft}
** For loc 106, MERS has amino acid L; DCLVS for L106C = 18
\end{flushleft}
\vspace{-2mm}
\caption{DCLVS values on insertion of Cys mutations on important AA of PLPro}
\label{DCLVS_nsp3}
\end{table}

\noindent
\emph{Conclusion from the results reported in Table~\ref{DCLVS_nsp3}:} In view of -ive DCLVS values, the mutations in Catalytic sites (C111, H272, D286) are not relevant for the CoV-2, probably because mutation on catalytic site will hinder PLPro cleavage function. For other important residues and hydrophobic cleft display higher DCLVS for CoV-2 (noted on $1^{st}$ row) compared to CoV (shown on $2^{nd}$ row). These results confirm higher efficiency of execution of PLPro tasks by CoV-2 compared to that of CoV.  Consequently, CoV-2 is expected to have higher viral load along with higher transmissibility.

\subsection{nsp5}
\label{CAML_nsp5_result}

\vspace{2mm}
As per the survey reported in Section~\ref{Task_nsp5}, active site with consensus cleavage region are H41 and C145.  The catalytic pocket is enclosed by two loop regions (AA 44 to 53 and AA 184 to 193) that play an important role in the catalysis. DCLVS values for H41, C145, and other important AAs (AA 44 to 53 and AA 184 to 193) are derived on introducing Cys mutation (Ala mutation if AA is Cys).                                                                                                                         DCLVS values for H41C and C145A are (-165 and 0) for CoV-2, while for CoV the corresponding DCLVS values are (615 and 0). The DCLVS values for AA (44 to 93) and (184 to 193), as reported in Table~\ref{DCLVS_nsp5}, display mostly -ive values for CoV-2, while CoV displays high +ive values.                                                                 The (-ive) values of CoV-2 AAs point to the fact that - no mutation is allowed at those sites probably because that will hinder nsp5 function (cleaving polyprotein at 11 sites). Consequently, as per Step 9 of the algorithm (reported in Section~\ref{ML_Framework_Design}), comparative evaluation in respect of transmissibility of CoV-2 and CoV nsp5 is derived based on the DCLVS values of CoV-2 and CoV wild nsps which are respectively 2932 and 2569, with CoV-2 value 14 \% higher. Hence, as per CAML model, due to higher CLVS value, higher viral load and transmissibility gets ensured for CoV-2 nsp5 compared to CoV for nsp5.

\begin{table}[h!]
\centering
\resizebox{\textwidth}{!}{  
\begin{tabular}{|c|c|c|c|c|c|c|c|c|c|c|c|c|c|c|c|c|c|c|c|c|}
\hline 
 & \multicolumn{20}{|c|}{DCLVS value for AA} \\ 
\hline 
 & C44 & T45 & S46 & E47 & D48 & M49 & L50 & N51 & P52 & N53 & P184 & F185 & V186 & D187 & R188 & Q189 & T190 & A191 & Q192 & A193 \\ 
\hline 
CoV-2 & -231 & -228 & -84 & -207 & 0 & -102 & 12 & -39 & -195 & -36 & -78 & -18 & -189 & 30 & 90 & -3 & -90 & 45 & -87 & 0  \\ 
\hline 
CoV & -231 & 342 & -84 & 573 & 0 & 678 & 12 & 241 & 585 & 744 & 458 & -18 & 327 & 30 & 90 & 513 & 426 & 45 & 489 & 0  \\ 
\hline 
\end{tabular} }
\caption{DCLVS values for AA (44 to 53) and (184 to 193); DCLVS values for H41C and C145A are (-165 and 0) and the corresponding values for CoV are (615 and 0)}
\label{DCLVS_nsp5}
\end{table}

\noindent
\emph{Observation from CAML model on MPro nsp5:} The AA locations of COV-2 displaying +ive values are identical to those of CoV in those locations. However, for most of the other AA locations, CoV-2 displays -ive DCLVS values, while CoV shows high +ive values. Being the main protease, mutations in catalytically important AAs are not relevant and probably not allowed by CoV-2 nsp5. On the other hand, CoV nsp5 permits such mutations leading to - probably inconsistent/wrong cleavage due to unwanted mutations permitted by CoV.

\subsection{nsp6}
\label{CAML_nsp6_result}

\vspace{2mm}
As per the survey reported in Section~\ref{Task_nsp6}, the authors \cite{miller2020coronavirus, choi2018autophagy} suggested that coronavirus may mimic host autophagy pathway for its replication. On executing Steps (3(a), 4, 5) of the algorithm reported in Section 5.2, we identify the 5 AA sequence of nsp6 RAA pairs. The list of conserved RAA pairs in CoV-2 and CoV in nsp6 AA chain are - (KR at loc 4-5 for CoV-2 and KK for CoV), (KHKH at location 61 to 64), and (RR at location 137 to 138). Table~\ref{DCLVS_nsp6} reports the DCLVS values for 5 AA around the RAA pairs along with DCLVS sum for three groups of 5 AAs. CAML model assumes that these 5 AA regions bind with host proteins for executing nsp6 task. For each group, CoV-2 DCLVS sum is greater than CoV.

\begin{table}[h!]
\centering
\resizebox{\textwidth}{!}{ 
\begin{tabular}{|c|c|c|c|c|c|c|c|c|c|c|c|c|c|c|c|c|c|c|}
\hline 
 & \multicolumn{6}{c|}{Group1 AA (loc 4 to 8)} & \multicolumn{6}{c|}{Group2 AA (loc 61 to 65)}  & \multicolumn{6}{c|}{Group3 AA (loc 136 to 140)}  \\ 
\hline 
 & K4 & R/K5 & T6 & I7 & K8 & Sum & K61 & H62 & K63 & H64 & A65 & Sum & A136 & R137 & R138 & V139 & W140 & Sum \\ 
\hline 
CoV-2 & 30 & 95 & 207 & 30 & 54 & 356 & 96 & 210 & 180 & 120 & 180 & 786 & 0 & 189 & 210 & 139 & 140 & 678 \\ 
\hline 
CoV & 18 & 30 & 75 & 15 & 0 & 75 & 48 & -279 & 120 & -279  & -309 & 195 & 189 & 189 & 0 & -351 & 95 & 473 \\ 
\hline 
\end{tabular} }
\caption{DCLVS values for three groups of 5 AAs around RAA pair along with their DCLVS sum}
\label{DCLVS_nsp6}
\end{table}

The results derived out of CAML model clearly demonstrates the higher efficiency of CoV-2 than CoV for availing host autophagosome pathways for viral replication through interaction with host proteins.

\subsection{nsp8}
\label{CAML_nsp8_result}

\vspace{2mm}
The nsp12-nsp8-nsp7 of RdRp complex with nsp8 and nsp7 as cofactors, plays a critical role for synthesis of RNA strand complimentary to RNA template. 
The effect of nsp8 mutations (reported in Table 2 of \cite{subissi2014one} for CoV) on interaction with nsp12 (with \% of interaction noted) are presented in Table~\ref{DCLVS_nsp8}.  The nsp8 mutants leading to significantly reduced interaction with nsp12 gets identified displaying high DCLVS values 210 and 147 with Cys mutations  (D99C, R190C).

\begin{table}[h!]
\centering
\resizebox{0.9\textwidth}{!}{ 
\begin{tabular}{|c|c|c|c|}
\hline 
\multirow{3}{*}{Mutant} & Results from Table 2 \cite{subissi2014one} & \multicolumn{2}{c|}{Results derived out of CAML model} \\  
 & \% Interaction with nsp12 & \multicolumn{2}{c|}{DCLVS Cys mutation} \\ \cline{3-4}
 &  & CoV-2 & CoV \\
\hline 
Wild & 100 +/- 19.5 & - & - \\ 
\hline 
D50A & 99.7 +/- 15.5 & -45 & -45 \\ 
\hline 
K58A & 90.7 +/- 15.5 & 0 & 0 \\ 
\hline 
K82A & 96.7 +/- 11.7 & 0 & 0 \\ 
\hline 
S85A & 105.2 +/- 9.7 & 30 & 30 \\ 
\hline 
D99A & 24.0 +/- 6.8 & 210 & 210 \\ 
\hline 
P116A & 20.6 +/- 3.9 & 30 & - 90 \\ 
\hline 
P183A & 15.8 +/- 5.2 & 0 & 15 \\ 
\hline 
R190A  & 19.4 +/- 12.6 & 147 & 147 \\ 
\hline 
\end{tabular} }
\caption{DCLVS values for CoV-2 and CoV for the mutation sites reported in \cite{subissi2014one}}
\label{DCLVS_nsp8}
\end{table}

The results derived out of CAML model establishes the fact that CoV-2 and CoV nsp8 mutants are similar in nature so far as their interaction with nsp12 is concerned.

\subsection{nsp10}
\label{CAML_nsp10_result}

\begin{figure}[h!]
\hfill
\subfigure[CoV-2 nsp10 wild CL graph \label{CL_Graph_nsp10_Cov2_wild}]{\includegraphics[width=13.3cm, height=4.7cm]{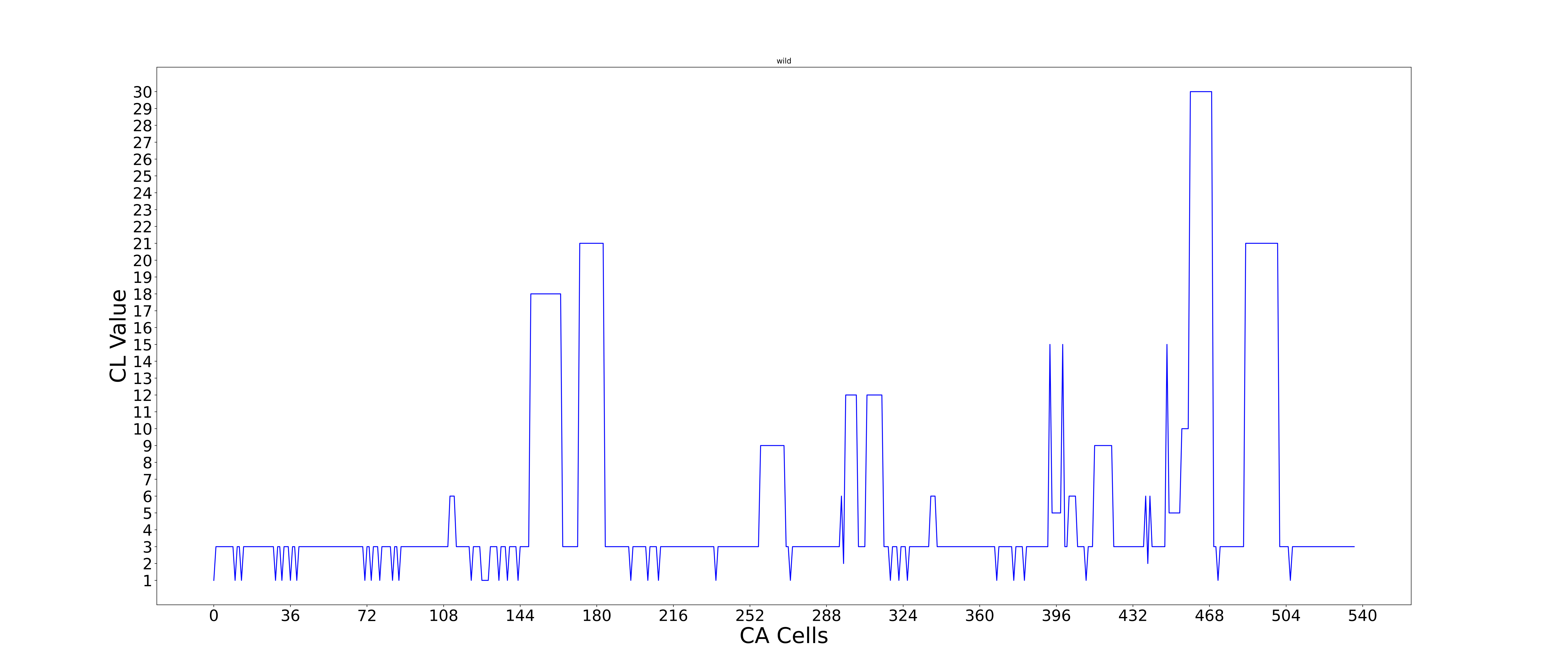}}
\hfill
\subfigure[CoV nsp10 wild CL graph\label{CL_Graph_nsp10_Cov_wild}]{\includegraphics[width=13.3cm, height=4.7cm]{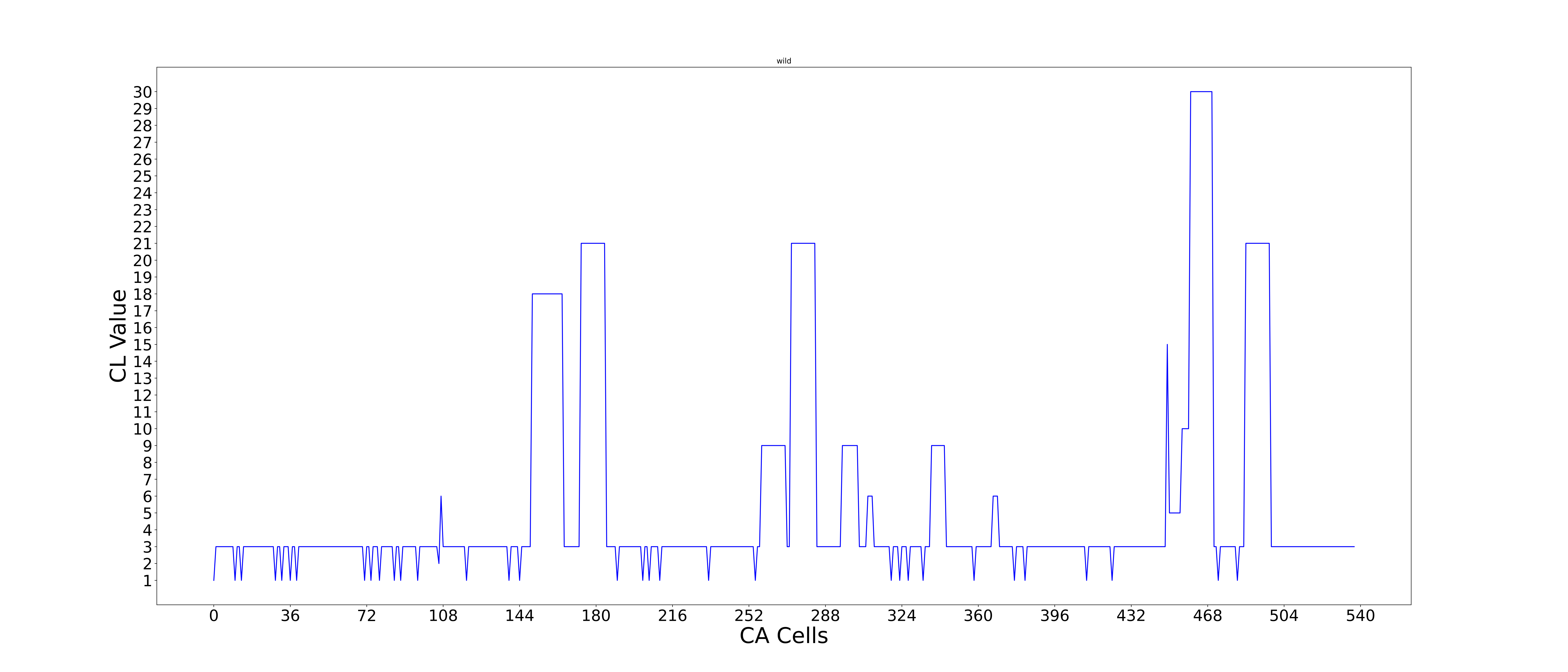}}
\hfill
\caption{Wild CL graphs for CoV-2 and CoV nsp10 (CL graphs are similar at NTD and CTD region and differ in the linker region between NTD and CTD) (Note - NTD and CTD respectively refer to start and end regions of AA chain of a protein)}
\end{figure}

\vspace{2mm}
The similarity of CoV-2 and CoV nsp10 AA sequences and their structures, as confirmed in \cite{rogstam2020crystal}, gets reflected by the similarity of CL signals graphs reported in Figure~\ref{CL_Graph_nsp10_Cov2_wild} and \ref{CL_Graph_nsp10_Cov_wild} in the NTD and CTD regions.                                                                                                                                                                   From the analysis of DCLVS values, as reported in Table~\ref{DCLVS_nsp10}, it can be observed that majority of mutants for SARS and MERS nsp10 are (-ive) valued and so not relevant. Hence as per Step 9 of ML Framework design algorithm (noted in Section~\ref{ML_Framework_Design}), we compare contribution of an nsp from comparative study of CLVS values of CoV-2 and CoV wild CL graphs. The CLVS values for CoV-2 and CoV are respectively 1630 and 1645 displaying difference of less than 1\%. Hence, we conclude that the contribution of cofactor nsp10 for interaction with nsp14 and nsp16 are identical.

\begin{table}[h!]
\centering
\resizebox{\textwidth}{!}{ 
\begin{tabular}{|c|c|c|c|c|c|c|c|c|c|c|}
\hline 
 & \multicolumn{4}{c|}{$1^{st}Zn$}  & \multicolumn{4}{c|}{$2^{nd}Zn$} & \multicolumn{2}{c|}{2 extra C} \\ \cline{2-11} 
 & Cys74A & Cys77A & His83C & Cys90A & Cys117A & Cys120A & Cys128A & Cys130A & C73A & C79A \\ 
\hline 
CoV-2 & 147 & 48 & -153 & 0 & -385 & -150 & -183 & -84 & 189 & -27 \\ 
\hline 
CoV & -105 & 48 & 207 & -63 & -385 & -150 & -144 & 0 & -63 & -27 \\ 
\hline 
MERS & -30 & 0 & -90 & -135 & -72 & -45 & 132 & -30 & -132 & -27 \\ 
\hline 
\end{tabular} }
\caption{DCLVS values for Zn binding sites; for most of the 19 mutants in Zn binding sites, the DCLVS values are (-ive) for CoV-2 and MERS}
\label{DCLVS_nsp10}
\end{table}

\subsection{nsp12 - RNA dependent RNA polymerase (RdRp) }
\label{CAML_nsp12_result}

\vspace{2mm}
From the survey of published literature reported in Section~\ref{survey_nsps}, the motifs and important AA are curated and noted in Table~\ref{DCLVS_nsp12}.  The 1st row of the table displays the AA for CoV-2, followed by 2nd row reporting DCLVS values derived on insertion of Cys mutations for each of the non-Cys  AA (Ala mutation for Cys AA) of row 1; similar results are reported on row (3, 4) for CoV, and row (5, 6) for MERS.

\begin{table}[h!]
\centering
\resizebox{\textwidth}{!}{ 
\begin{tabular}{|c|c|c|c|c|c|c|c|c|c|c|c|c|c|c|c|c|c|c|c|}
\hline 
Virus & \multicolumn{3}{c|}{Motif} & \multicolumn{6}{c|}{Active Sites} & \multicolumn{5}{c|}{\textbf{Fe-S Catalytic site}} & \multicolumn{5}{c|}{Fe-S Interface} \\ 
\hline 
\textbf{CoV-2} & L731 & Y732 & R733 & S759 & D760 & D761 & K798 & S814 & Sum & C487 & H642 & C645 & C646 & Sum & H295 & C301 & C306 & C310 & Sum  \\ 
\hline 
DCLVS & 240 & 0 & 12 & 0 & (-45) & 30 & 174 & 0 & 174 & 0 & 596 & 72 & (-75) & 668 & 375 & 0 & (-189) & 0 & 375  \\ 
\hline 
\textbf{CoV} & V727 & Y728 & R729 & S755 & D756 & D757 & K794 & D810  & - & C483 & H638 & C641 & C642 & - & H291 & C297 & C302 & C306  & -  \\ 
\hline 
DCLVS & 573 & 30 & 24 & 45 & 45 & 18 & 174 & 0 & 174 & 0 & 300 & 147 & 111 & 558 & 342 & (-72) & (-144) & 0 & 342  \\ 
\hline 
\textbf{MERS} & L728 & Y729 & V730 & S760 & D761 & D762 & K799 & S815 & - & C488 & H643 & C646 & C647 & - & H296 & C302 & C307 & C311 & -  \\ 
\hline 
DCLVS & 187 & 234 & (-18) & 0 & (-45) & 123 & 174 & (-18) & 297 & 0  & 1473 & (-83) & (-287) & 1473 & (-82) & 0 & 0 & 0 & 0  \\ 
\hline 
\end{tabular} }
\caption{DCLVS values for mutants on active sites retrieved from published literature \cite{maio2021fe, gao2020structure} noted on rows 2, 4, 6 for CoV-2, CoV, MERS nsp12 respectively. Rows 3, 5, 7 reports the DCLVS values derived out of Cys mutations for AA other than Cys AA and Ala mutations for Cys AA in the list}
\label{DCLVS_nsp12}
\end{table}

Out of the four group of AAs (Motif, Active sites, Fe-S Catalytic sites, Fe-S Interface) shown in Table~\ref{DCLVS_nsp12}, the third group, Fe-S catalytic domain (shown in bold letter) of nsp12, is the most crucial element for RNA polymerase activity of nsp12 - nsp8 - nsp7 complex \cite{maio2021fe}. The DCLVS values in the Fe-S catalytic site displays higher value of 668 for CoV-2 compared to 558 for CoV. Higher DCLVS of CoV-2 than CoV confirms execution of nsp12 tasks with higher efficiency. Consequently, as per CAML model, viral Load and transmissibility of CoV-2 will be higher than that of CoV by approximately 20 \%. On the other hand MERS with DCLVS value 1473 stands apart for SARS (CoV2, and CoV).

\subsection{nsp14 - Exoribonuclease}
\label{CAML_nsp14_result}

\vspace{2mm}
The Enzymatic activity of the nsp14 Exoribonuclease (coordinated by metal ions) is reviewed in Section~\ref{Task_nsp14} of survey in Section~\ref{survey_nsps}. The activities are executed by the five catalytic DEEDH amino acids (D90, E92, E191, D273, H268). In order to assess the importance of each of the catalytic AA, Cys mutation is introduced followed by evaluation of DCLVS values for each mutant. Higher DCLVS value, as reported in Table~\ref{DCLCS_nsp14}, points to higher efficiency of execution of the catalytic function. As per this evaluation, CoV-2 is a clear winner compared to CoV and MERS so far as execution of the catalytic function of nsp14 is concerned.

\begin{table}[h!]
\centering
\resizebox{\textwidth}{!}{ 
\begin{tabular}{|c|c|c|c|c|c|c|}
\hline 
\multicolumn{7}{|c|}{CAML DCLVS value for the catalytic amino acids : (CLVS wild : CoV-2=2358, CoV=2854, MERS=2567)} \\ 
\hline 
 & D90C & E92C & E191C & D273C & H268C & Sum \\ 
\hline 
CoV-2 & 55 & 332 & 206 & 165 & 214 & 972 \\ 
\hline 
CoV & 55 & -225 & -351 & 165 & -502 & 220 \\ 
\hline 
MERS & -45 & -111 & -319 & 0 & -51 &  \\ 
\hline 
\end{tabular} }
\caption{DCLVS values for important AAs of nsp14; CoV-2 DCLVS values are equal or higher compared to CoV and MERS}
\label{DCLCS_nsp14}
\end{table}

The results derived out of CAML model establish the highest efficiency of execution of nsp14 tasks leading to highest transmissibility for CoV-2 among three viruses.

\subsection{nsp15 - Endoribonuclease (EndoU)}
\label{CAML_nsp15_result}

\vspace{2mm}
As reviewed in Section~\ref{Task_nsp15} (of survey reported in Section~\ref{survey_nsps}), six important amino acids associated with EndoU catalytic activity are (H234, H249, K289 , S294, Y343, L346); the corresponding AA for MERS are (H231, H246, K285,   - ,   Y339, R341).  In order to assess the importance of each of the six AAs for nsp15 function, CAML model introduces Cys mutation and evaluate the DCLVS of CoV-2 nsp15 and CoV nsp15. The DCLVS values for six AAs and their Sum are reported in Table~\ref{DCLVS_nsp15} column 1 to 7. The difference of 5.4 \% between COV-2 and COV DCLVS sum shown on column 8. CoV-2 wild (918) and CoV nsp15 wild (1251) with their difference as 333 are presented in column 9. On the other hand, the DCLVS for CoV-2 and CoV for the region covering AA location 100 to 200 (the middle domain of nsp15) is reported in column 10. High difference of values 434 in this domain confirms the result reported in \cite{kim2020crystal}.

\begin{table}[h!]
\centering
\resizebox{\textwidth}{!}{ 
\begin{tabular}{|c|c|c|c|c|c|c|c|c|c|c|}
\hline 
Col & 1 & 2 & 3 & 4 & 5 & 6 & 7 & 8 & 9 & 10 (middle domain) \\ 
\hline 
Covid & H234C & H249C & K289C & V294 & Y342C & Q346C & Sum & \% diff (CoV-2, CoV) & DCLVS - wild & DCLVS diff (AA loc 100 to 200)  \\ 
\hline 
CoV-2 & 366 & 375 & 180 & 192 & 0 & 147 & 1260 & - & - & - \\ 
\hline 
CoV & 402 & 411 & 180 & 192 & 0 & 147 & 1332 & 5.4 \%  & CoV-2 (918), CoV (1251) & 434 \\ 
\hline 
 & H231 & H246 & K285C & - & Y339C & R341C & &  &  &  \\ 
\hline 
MERS & 168 & 54 & 252  & • & 0 & 84 & 558 &  &  &  \\ 
\hline 
\end{tabular} }
\caption{DCLVS values of nsp15 CoV-2, CoV, MERS AAs associated with nsp15 catalytic function. Column 8 reports the percentage difference of DCLVS SUM as 5.4 \% between CoV-2 and CoV nsp15. Column 9  reports difference between wild CoV-2 and CoV for the full length wild CL graphs, while column 10 shows the difference of DCLVS for  the middle domain (AA loc 100 to 200)}
\label{DCLVS_nsp15}
\end{table}

As per the results reported in Table~\ref{DCLVS_nsp15}, the DCLVS Sum are similar with percentage difference of 5.4\% (Col 8, Table~\ref{DCLVS_nsp15}).  In view of similar degree of host immune evasion for CoV-2 and CoV EndoU function, we conclude that transmissibility for CoV-2 and CoV will be of same order so far as execution of task by nsp15 is concerned.

\subsection{nsp16 - 2'- O - methyltransferase}
\label{CAML_nsp16_result}

\vspace{2mm}
As reviewed in Section~\ref{Task_nsp16} of survey reported in Section~\ref{survey_nsps}, the binding site AA (K46, D130, K170, and E203) - the KDKE motif of nsp16, are conserved among CoV-2, CoV, MERS in order to down regulate host immune response.                                                                                                                                                                                                                                                                                     In order to evaluate the importance of these AA for execution of nsp16 task, CAML model introduces Cys mutation. The DCLVS values (Table~\ref{DCLVS_nsp16}) for CoV-2 is higher than or equal to that of CoV and MERS. These results point to highest efficiency for execution of nsp16 function of CoV-2 compared to CoV and MERS. Consequently, downregulation of host immune response is higher for CoV-2 compared to CoV leading to higher transmissibility of CoV-2 compared to that of CoV.

\begin{table}[h!]
\centering
\resizebox{0.7\textwidth}{!}{ 
\begin{tabular}{|c|c|c|c|c|c|}
\hline 
 & K46C & D130C & K170C & E203C & Sum \\ 
\hline 
CoV-2 & 189 & 168 & 294 & 147 & 798 \\ 
\hline 
CoV & 189 & 168 & 30 & -18 & 357 \\ 
\hline 
MERS & 189 & 168 & 189 & -252 & 546 \\ 
\hline 
\end{tabular} }
\caption{DCLVS values of four AAs important for execution of nsp16 task showing equal or higher DCLVS values for CoV-2 compared to CoV (Note - DCLVS values GE 50 are only considered)}
\label{DCLVS_nsp16}
\end{table}

Further, the nsp10 AAs (L45, H80, K93, G94, Y96 (F96 for MERS) interacts respectively with nsp16 AAs (Q87, D102, S105, D106, Q87 (R87 for MERS)) in nsp10-nsp16 complex to stabilize the binding pocket of nsp16. So, we evaluate DCLVS values for each of these AA of nsp16 on inserting Cys mutation. Table~\ref{DCLVS_nsp16_2} displays higher DCLVS Sum for CoV-2 nsp16 compared to CoV. These results confirm higher degree of stabilization of the nsp16-nsp10 CoV-2 complex compared to CoV for execution of nsp16 task.

\begin{table}[h!]
\centering
\resizebox{\textwidth}{!}{ 
\begin{tabular}{|c|c|c|c|c|c|c|c|c|c|c|c|c|}
\hline 
 & \multicolumn{6}{|c|}{nsp16  AAs} & \multicolumn{6}{|c|}{nsp10 AAs (derived from nsp10 mutation file)} \\ 
\hline 
 & Q87 & D102 & S105 & D106 & Q87 & Sum & L45 & H80 & K93 & G94 & Y/F96 & Sum \\ 
\hline 
CoV-2 & 126 & 0 & 45 & 189 & 126 & 486 & 36 & -69 & 0 & 135 & 105 & 240 \\ 
\hline 
CoV & -219 & 0 & 45 & 189 & -219 & 234 & 36 & 60 & 180 & 0  & 0  & 240 \\ 
\hline 
MERS & 6 & 78 & 45 & 189 & 6 & 267 & 234 & -183 & 0 & 135 & 105 & 474 \\ 
\hline 
\end{tabular} }
\caption{DCLVS values for the AAs of nsp16 - nsp10 complex; Nsp10 AAs (L45, H80, K93, G94, Y96 (F96 for MERS) interacts respectively with nsp16 AAs (Q87, D102, S105, D106, Q87 (R87 for MERS). (Note - DCLVS values GE 50 are only considered; Y96 for COV-2 and CoV and F96 for MERS)}
\label{DCLVS_nsp16_2}
\end{table}

Higher DCLVS values of CoV-2 along with higher stability of of nsp16-nsp10 complex, point to higher efficiency of execution of nsp16 function. As a result, downregulation of host immune response by CoV-2 will be higher compared to CoV. Consequently, CoV-2 will have higher replication and transmissibility compared to CoV.

\subsection{Summary of results derived out of CAML model for nsps other than nsp1}

\vspace{2mm}
The model confirms higher contribution of CoV-2 nsp3, nsp5, nsp6, nsp12, nsp14, nsp16 compared to those of CoV. On the other hand, contribution of Cov-2 and CoV have been observed to be similar in nature for nsp8, nsp10, nsp15.

\section{Conclusion}

Significantly higher transmissibility of SARS CoV-2 (2019) compared to SARS CoV (2003) is an established fact. The reason for CoV-2 higher transmissibility can be attributed to mutations of structural proteins (Spike S, Nucleocapsid N, Membrane M, and Envelope E) and the role played by non-structural proteins (nsps) and accessory proteins (ORFs) for viral replication, assembly, and shedding. Based on CAML model, in \cite{hazari2022analysis} we reported mutational study to compare the transmissibility factor of CoV-2 and CoV envelop proteins.  The current paper compares the                                                                                                                        transmissibility of CoV-2 and CoV nsps to ascertain their contribution towards higher transmissibility of CoV-2. The nsp1 is the major determinant factor subsequent to viral entry in host for - (i) shutting off host translation, while ensuring (ii) viral translation, and (iii) downregulation of host immune response. Execution of these three tasks (marked as Ta1, Ta2, and Ta3) by CoV-2 and CoV are compared to establish that CoV-2 nsp1 contributes towards higher transmissibility due to higher efficiency of host translational shut off compared to that of CoV. Cov-2 also adversely affects host immune defense with higher efficiency than CoV.  The numerical figures derived out of CAML model for nsp1 point to higher transmissibility of CoV-2 compared to CoV.  Further, contribution of other major nsps (nsp3, nsp5, nsp6, nsp8, nsp10, nsp12, nsp14, nsp15, nsp16) towards higher transmissibility of CoV-2 compared to CoV is established from CAML model.

Our next work will report CAML model for mutational study on Spike protein sequence. This model will report a list of `Neutralizing Antibody Escape' mutants out of the Mutational Hotspots predicted in spike protein AA sequence. We expect the quantitative analysis reported out of CAML model will provide a platform for design of robust vaccine and therapeutic agents in the context of appearance new CoV-2 strains. Further, as many experts predicted, due to continuous environmental degradation, new types of viruses are likely to appear in the near future. CAML platform can help to fight next pandemic situation for quick design of vaccines and therapeutic agents.

\section*{References}
\bibliographystyle{elsarticle-num} 
\bibliography{References}

\begin{thebibliography}{10}
\expandafter\ifx\csname url\endcsname\relax
  \def\url#1{\texttt{#1}}\fi
\expandafter\ifx\csname urlprefix\endcsname\relax\def\urlprefix{URL }\fi
\expandafter\ifx\csname href\endcsname\relax
  \def\href#1#2{#2} \def\path#1{#1}\fi

\bibitem{roossinck2017symbiosis}
M.~J. Roossinck, E.~R. Baz{\'a}n, Symbiosis: viruses as intimate partners,
  Annual Review of Virology 4 (2017) 123--139.

\bibitem{yamamoto2021human}
S.~Yamamoto, M.~Saito, A.~Tamura, D.~Prawisuda, T.~Mizutani, H.~Yotsuyanagi,
  The human microbiome and covid-19: A systematic review, PloS one 16~(6)
  (2021) e0253293.

\bibitem{shen2021lysine}
Z.~Shen, G.~Zhang, Y.~Yang, M.~Li, S.~Yang, G.~Peng, Lysine 164 is critical for
  sars-cov-2 nsp1 inhibition of host gene expression, The Journal of General
  Virology 102~(1).

\bibitem{schubert2020sars}
K.~Schubert, E.~D. Karousis, A.~Jomaa, A.~Scaiola, B.~Echeverria, L.-A.
  Gurzeler, M.~Leibundgut, V.~Thiel, O.~M{\"u}hlemann, N.~Ban, Sars-cov-2 nsp1
  binds the ribosomal mrna channel to inhibit translation, Nature structural \&
  molecular biology 27~(10) (2020) 959--966.

\bibitem{nakagawa2021mechanisms}
K.~Nakagawa, S.~Makino, Mechanisms of coronavirus nsp1-mediated control of host
  and viral gene expression, Cells 10~(2) (2021) 300.

\bibitem{simeoni2021nsp1}
M.~Simeoni, T.~Cavinato, D.~Rodriguez, D.~Gatfield, I (nsp1)ecting
  sars-cov-2--ribosome interactions, Communications biology 4~(1) (2021) 1--5.

\bibitem{terada2017mers}
Y.~Terada, K.~Kawachi, Y.~Matsuura, W.~Kamitani, Mers coronavirus nsp1
  participates in an efficient propagation through a specific interaction with
  viral rna, Virology 511 (2017) 95--105.

\bibitem{benedetti2020emerging}
F.~Benedetti, G.~A. Snyder, M.~Giovanetti, S.~Angeletti, R.~C. Gallo,
  M.~Ciccozzi, D.~Zella, Emerging of a sars-cov-2 viral strain with a deletion
  in nsp1, Journal of Translational Medicine 18~(329) (2020) 1--6.

\bibitem{clark2021structure}
L.~K. Clark, T.~J. Green, C.~M. Petit, Structure of nonstructural protein 1
  from sars-cov-2, Journal of Virology 95~(4) (2021) e02019--20.

\bibitem{kumar2020sars}
A.~Kumar, A.~Kumar, P.~Kumar, N.~Garg, R.~Giri, Sars-cov-2 nsp1 c-terminal
  region (residues 130-180) is an intrinsically disordered region, BioRxiv.

\bibitem{tidu2021viral}
A.~Tidu, A.~Janvier, L.~Schaeffer, P.~Sosnowski, L.~Kuhn, P.~Hammann,
  E.~Westhof, G.~Eriani, F.~Martin, The viral protein nsp1 acts as a ribosome
  gatekeeper for shutting down host translation and fostering sars-cov-2
  translation, RNA 27~(3) (2021) 253--264.

\bibitem{vankadari2020structure}
N.~Vankadari, N.~N. Jeyasankar, W.~J. Lopes, Structure of the sars-cov-2
  nsp1/5'-untranslated region complex and implications for potential
  therapeutic targets, a vaccine, and virulence, The journal of physical
  chemistry letters 11~(22) (2020) 9659--9668.

\bibitem{de2020translational}
S.~de~Breyne, C.~Vindry, O.~Guillin, L.~Cond{\'e}, F.~Mure, H.~Gruffat,
  L.~Chavatte, T.~Ohlmann, Translational control of coronaviruses, Nucleic
  Acids Research 48~(22) (2020) 12502--12522.

\bibitem{miao2021secondary}
Z.~Miao, A.~Tidu, G.~Eriani, F.~Martin, Secondary structure of the sars-cov-2
  5'-utr, RNA biology 18~(4) (2021) 447--456.

\bibitem{tanaka2012severe}
T.~Tanaka, W.~Kamitani, M.~L. DeDiego, L.~Enjuanes, Y.~Matsuura, Severe acute
  respiratory syndrome coronavirus nsp1 facilitates efficient propagation in
  cells through a specific translational shutoff of host mrna, Journal of
  virology 86~(20) (2012) 11128--11137.

\bibitem{yang2015structure}
D.~Yang, J.~L. Leibowitz, The structure and functions of coronavirus genomic 3'
  and 5' ends, Virus research 206 (2015) 120--133.

\bibitem{mohammadi2021transcription}
M.~Mohammadi-Dehcheshmeh, S.~M. Moghbeli, S.~Rahimirad, I.~O. Alanazi, Z.~S.~A.
  Shehri, E.~Ebrahimie, A transcription regulatory sequence in the 5'
  untranslated region of sars-cov-2 is vital for virus replication with an
  altered evolutionary pattern against human inhibitory micrornas, Cells 10~(2)
  (2021) 319.

\bibitem{kikkert2020innate}
M.~Kikkert, Innate immune evasion by human respiratory rna viruses, Journal of
  innate immunity 12~(1) (2020) 4--20.

\bibitem{thoms2020structural}
M.~Thoms, R.~Buschauer, M.~Ameismeier, L.~Koepke, T.~Denk, M.~Hirschenberger,
  H.~Kratzat, M.~Hayn, T.~Mackens-Kiani, J.~Cheng, et~al., Structural basis for
  translational shutdown and immune evasion by the nsp1 protein of sars-cov-2,
  Science 369 (2020) 1249--1255.

\bibitem{taefehshokr2020covid}
N.~Taefehshokr, S.~Taefehshokr, N.~Hemmat, B.~Heit, Covid-19: perspectives on
  innate immune evasion, Frontiers in immunology (2020) 580641.

\bibitem{lokugamage2015middle}
K.~G. Lokugamage, K.~Narayanan, K.~Nakagawa, K.~Terasaki, S.~I. Ramirez,
  C.-T.~K. Tseng, S.~Makino, Middle east respiratory syndrome coronavirus nsp1
  inhibits host gene expression by selectively targeting mrnas transcribed in
  the nucleus while sparing mrnas of cytoplasmic origin, Journal of virology
  89~(21) (2015) 10970--10981.

\bibitem{lei2018nsp3}
J.~Lei, Y.~Kusov, R.~Hilgenfeld, Nsp3 of coronaviruses: Structures and
  functions of a large multi-domain protein, Antiviral research 149 (2018)
  58--74.

\bibitem{shin2020papain}
D.~Shin, R.~Mukherjee, D.~Grewe, D.~Bojkova, K.~Baek, A.~Bhattacharya,
  L.~Schulz, M.~Widera, A.~R. Mehdipour, G.~Tascher, et~al., Papain-like
  protease regulates sars-cov-2 viral spread and innate immunity, Nature 587
  (2020) 657--662.

\bibitem{stasiulewicz2021sars}
A.~Stasiulewicz, A.~W. Maksymiuk, M.~L. Nguyen, B.~Be{\l}za, J.~I. Sulkowska,
  Sars-cov-2 papain-like protease potential inhibitors—in silico quantitative
  assessment, International journal of molecular sciences 22 (2021) 3957.

\bibitem{osipiuk2021structure}
J.~Osipiuk, S.-A. Azizi, S.~Dvorkin, M.~Endres, R.~Jedrzejczak, K.~A. Jones,
  S.~Kang, R.~S. Kathayat, Y.~Kim, V.~G. Lisnyak, et~al., Structure of
  papain-like protease from sars-cov-2 and its complexes with non-covalent
  inhibitors, Nature communications 12~(743) (2021) 1--9.

\bibitem{armstrong2021biochemical}
L.~A. Armstrong, S.~M. Lange, V.~Dee~Cesare, S.~P. Matthews, R.~S. Nirujogi,
  I.~Cole, A.~Hope, F.~Cunningham, R.~Toth, R.~Mukherjee, et~al., Biochemical
  characterization of protease activity of nsp3 from sars-cov-2 and its
  inhibition by nanobodies, PloS one 16~(7) (2021) e0253364.

\bibitem{gao2021crystal}
X.~Gao, B.~Qin, P.~Chen, K.~Zhu, P.~Hou, J.~A. Wojdyla, M.~Wang, S.~Cui,
  Crystal structure of sars-cov-2 papain-like protease, Acta Pharmaceutica
  Sinica B 11~(1) (2021) 237--245.

\bibitem{bosken2020insights}
Y.~K. Bosken, T.~Cholko, Y.-C. Lou, K.-P. Wu, C.-e.~A. Chang, Insights into
  dynamics of inhibitor and ubiquitin-like protein binding in sars-cov-2
  papain-like protease, Frontiers in molecular biosciences (2020) 174.

\bibitem{baez2015sars}
Y.~M. B{\'a}ez-Santos, S.~E.~S. John, A.~D. Mesecar, The sars-coronavirus
  papain-like protease: structure, function and inhibition by designed
  antiviral compounds, Antiviral research 115 (2015) 21--38.

\bibitem{shen2021potent}
Z.~Shen, K.~Ratia, L.~Cooper, D.~Kong, H.~Lee, Y.~Kwon, Y.~Li, S.~Alqarni,
  F.~Huang, O.~Dubrovskyi, et~al., Potent, novel sars-cov-2 plpro inhibitors
  block viral replication in monkey and human cell cultures, BioRxiv.

\bibitem{citarella2021sars}
A.~Citarella, A.~Scala, A.~Piperno, N.~Micale, Sars-cov-2 mpro: A potential
  target for peptidomimetics and small-molecule inhibitors, Biomolecules 11
  (2021) 607.

\bibitem{grottesi2020computational}
A.~Grottesi, N.~Be{\v{s}}ker, A.~Emerson, C.~Manelfi, A.~R. Beccari,
  F.~Frigerio, E.~Lindahl, C.~Cerchia, C.~Talarico, Computational studies of
  sars-cov-2 3clpro: Insights from md simulations, International journal of
  molecular sciences 21 (2020) 5346.

\bibitem{ferreira2020biochemical}
J.~C. Ferreira, W.~M. Rabeh, Biochemical and biophysical characterization of
  the main protease, 3-chymotrypsin-like protease (3clpro) from the novel
  coronavirus sars-cov 2, Scientific reports 10~(22200) (2020) 1--10.

\bibitem{gossen2021blueprint}
J.~Gossen, S.~Albani, A.~Hanke, B.~P. Joseph, C.~Bergh, M.~Kuzikov,
  E.~Costanzi, C.~Manelfi, P.~Storici, P.~Gribbon, et~al., A blueprint for high
  affinity sars-cov-2 mpro inhibitors from activity-based compound library
  screening guided by analysis of protein dynamics, ACS pharmacology \&
  translational science 4 (2021) 1079--1095.

\bibitem{lee2020crystallographic}
J.~Lee, L.~J. Worrall, M.~Vuckovic, F.~I. Rosell, F.~Gentile, A.-T. Ton, N.~A.
  Caveney, F.~Ban, A.~Cherkasov, M.~Paetzel, et~al., Crystallographic structure
  of wild-type sars-cov-2 main protease acyl-enzyme intermediate with
  physiological c-terminal autoprocessing site, Nature communications 11~(5877)
  (2020) 1--9.

\bibitem{miczi2020identification}
M.~Miczi, M.~Golda, B.~Kunkli, T.~Nagy, J.~T{\H{o}}zs{\'e}r, J.~A.
  M{\'o}ty{\'a}n, Identification of host cellular protein substrates of
  sars-cov-2 main protease, International journal of molecular sciences 21
  (2020) 9523.

\bibitem{jin2020structure}
Z.~Jin, X.~Du, Y.~Xu, Y.~Deng, M.~Liu, Y.~Zhao, B.~Zhang, X.~Li, L.~Zhang,
  C.~Peng, et~al., Structure of mpro from sars-cov-2 and discovery of its
  inhibitors, Nature 582 (2020) 289--293.

\bibitem{roe2021targeting}
M.~K. Roe, N.~A. Junod, A.~R. Young, D.~C. Beachboard, C.~C. Stobart, Targeting
  novel structural and functional features of coronavirus protease nsp5
  (3clpro, mpro) in the age of covid-19, The Journal of general virology
  102~(001558) (2021) 1--16.

\bibitem{cottam2014coronavirus}
E.~M. Cottam, M.~C. Whelband, T.~Wileman, Coronavirus nsp6 restricts
  autophagosome expansion, Autophagy 10~(8) (2014) 1426--1441.

\bibitem{benvenuto2020evolutionary}
D.~Benvenuto, S.~Angeletti, M.~Giovanetti, M.~Bianchi, S.~Pascarella, R.~Cauda,
  M.~Ciccozzi, A.~Cassone, Evolutionary analysis of sars-cov-2: how mutation of
  non-structural protein 6 (nsp6) could affect viral autophagy, Journal of
  Infection 81 (2020) e24--e27.

\bibitem{miller2020coronavirus}
K.~Miller, M.~E. McGrath, Z.~Hu, S.~Ariannejad, S.~Weston, M.~Frieman, W.~T.
  Jackson, Coronavirus interactions with the cellular autophagy machinery,
  Autophagy 16~(12) (2020) 2131--2139.

\bibitem{choi2018autophagy}
Y.~Choi, J.~W. Bowman, J.~U. Jung, Autophagy during viral infection—a
  double-edged sword, Nature Reviews Microbiology 16 (2018) 341--354.

\bibitem{santerre2021sars}
M.~Santerre, S.~P. Arjona, C.~N. Allen, N.~Shcherbik, B.~E. Sawaya, Why do
  sars-cov-2 nsps rush to the er?, Journal of neurology 268 (2021) 2013--2022.

\bibitem{schmidt2021sars}
N.~Schmidt, C.~A. Lareau, H.~Keshishian, S.~Ganskih, C.~Schneider, T.~Hennig,
  R.~Melanson, S.~Werner, Y.~Wei, M.~Zimmer, et~al., The sars-cov-2
  rna--protein interactome in infected human cells, Nature microbiology 6
  (2021) 339--353.

\bibitem{steele2015role}
S.~Steele, J.~Brunton, T.~Kawula, The role of autophagy in intracellular
  pathogen nutrient acquisition, Frontiers in cellular and infection
  microbiology 5 (2015) 51.

\bibitem{snijder2020unifying}
E.~J. Snijder, R.~W. Limpens, A.~H. de~Wilde, A.~W. de~Jong, J.~C.
  Zevenhoven-Dobbe, H.~J. Maier, F.~F. Faas, A.~J. Koster, M.~B{\'a}rcena, A
  unifying structural and functional model of the coronavirus replication
  organelle: Tracking down rna synthesis, PLoS biology 18~(6) (2020) e3000715.

\bibitem{hillen2020structure}
H.~S. Hillen, G.~Kokic, L.~Farnung, C.~Dienemann, D.~Tegunov, P.~Cramer,
  Structure of replicating sars-cov-2 polymerase, Nature 584 (2020) 154--156.

\bibitem{subissi2014one}
L.~Subissi, C.~C. Posthuma, A.~Collet, J.~C. Zevenhoven-Dobbe, A.~E.
  Gorbalenya, E.~Decroly, E.~J. Snijder, B.~Canard, I.~Imbert, One severe acute
  respiratory syndrome coronavirus protein complex integrates processive rna
  polymerase and exonuclease activities, Proceedings of the National Academy of
  Sciences 111~(37) (2014) E3900--E3909.

\bibitem{biswal2021two}
M.~Biswal, S.~Diggs, D.~Xu, N.~Khudaverdyan, J.~Lu, J.~Fang, G.~Blaha, R.~Hai,
  J.~Song, Two conserved oligomer interfaces of nsp7 and nsp8 underpin the
  dynamic assembly of sars-cov-2 rdrp, Nucleic acids research 49~(10) (2021)
  5956--5966.

\bibitem{reshamwala2021mutations}
S.~M. Reshamwala, V.~Likhite, M.~S. Degani, S.~S. Deb, S.~B. Noronha, Mutations
  in sars-cov-2 nsp7 and nsp8 proteins and their predicted impact on
  replication/transcription complex structure, Journal of medical virology
  93~(7) (2021) 4616--4619.

\bibitem{rogstam2020crystal}
A.~Rogstam, M.~Nyblom, S.~Christensen, C.~Sele, V.~O. Talibov, T.~Lindvall,
  A.~A. Rasmussen, I.~Andr{\'e}, Z.~Fisher, W.~Knecht, et~al., Crystal
  structure of non-structural protein 10 from severe acute respiratory syndrome
  coronavirus-2, International journal of molecular sciences 21~(19) (2020)
  7375.

\bibitem{kasuga2021innate}
Y.~Kasuga, B.~Zhu, K.-J. Jang, J.-S. Yoo, Innate immune sensing of coronavirus
  and viral evasion strategies, Experimental \& molecular medicine 53~(5)
  (2021) 723--736.

\bibitem{bouvet2014coronavirus}
M.~Bouvet, A.~Lugari, C.~C. Posthuma, J.~C. Zevenhoven, S.~Bernard, S.~Betzi,
  I.~Imbert, B.~Canard, J.-C. Guillemot, P.~L{\'e}cine, et~al., Coronavirus
  nsp10, a critical co-factor for activation of multiple replicative enzymes,
  Journal of Biological Chemistry 289~(37) (2014) 25783--25796.

\bibitem{saramago2021new}
M.~Saramago, C.~B{\'a}rria, V.~G. Costa, C.~S. Souza, S.~C. Viegas,
  S.~Domingues, D.~Lousa, C.~M. Soares, C.~M. Arraiano, R.~G. Matos, New
  targets for drug design: importance of nsp14/nsp10 complex formation for the
  3'-5'exoribonucleolytic activity on sars-cov-2, The FEBS journal.

\bibitem{krafcikova2020structural}
P.~Krafcikova, J.~Silhan, R.~Nencka, E.~Boura, Structural analysis of the
  sars-cov-2 methyltransferase complex involved in rna cap creation bound to
  sinefungin, Nature communications 11~(3717) (2020) 1--7.

\bibitem{lin2021crystal}
S.~Lin, H.~Chen, Z.~Chen, F.~Yang, F.~Ye, Y.~Zheng, J.~Yang, X.~Lin, H.~Sun,
  L.~Wang, et~al., Crystal structure of sars-cov-2 nsp10 bound to nsp14-exon
  domain reveals an exoribonuclease with both structural and functional
  integrity, Nucleic acids research 49~(9) (2021) 5382--5392.

\bibitem{naydenova2021structure}
K.~Naydenova, K.~W. Muir, L.-F. Wu, Z.~Zhang, F.~Coscia, M.~J. Peet,
  P.~Castro-Hartmann, P.~Qian, K.~Sader, K.~Dent, et~al., Structure of the
  sars-cov-2 rna-dependent rna polymerase in the presence of favipiravir-rtp,
  Proceedings of the National Academy of Sciences 118~(7).

\bibitem{kirchdoerfer2019structure}
R.~N. Kirchdoerfer, A.~B. Ward, Structure of the sars-cov nsp12 polymerase
  bound to nsp7 and nsp8 co-factors, Nature communications 10~(2342) (2019)
  1--9.

\bibitem{yan2021cryo}
L.~Yan, J.~Ge, L.~Zheng, Y.~Zhang, Y.~Gao, T.~Wang, Y.~Huang, Y.~Yang, S.~Gao,
  M.~Li, et~al., Cryo-em structure of an extended sars-cov-2 replication and
  transcription complex reveals an intermediate state in cap synthesis, Cell
  184~(1) (2021) 184--193.

\bibitem{maio2021fe}
N.~Maio, B.~A. Lafont, D.~Sil, Y.~Li, J.~M. Bollinger~Jr, C.~Krebs, T.~C.
  Pierson, W.~M. Linehan, T.~A. Rouault, Fe-s cofactors in the sars-cov-2
  rna-dependent rna polymerase are potential antiviral targets, Science 373
  (2021) 236--241.

\bibitem{peng2020structural}
Q.~Peng, R.~Peng, B.~Yuan, J.~Zhao, M.~Wang, X.~Wang, Q.~Wang, Y.~Sun, Z.~Fan,
  J.~Qi, et~al., Structural and biochemical characterization of the
  nsp12-nsp7-nsp8 core polymerase complex from sars-cov-2, Cell reports 31~(11)
  (2020) 107774.

\bibitem{aftab2020analysis}
S.~O. Aftab, M.~Z. Ghouri, M.~U. Masood, Z.~Haider, Z.~Khan, A.~Ahmad,
  N.~Munawar, Analysis of sars-cov-2 rna-dependent rna polymerase as a
  potential therapeutic drug target using a computational approach, Journal of
  translational medicine 18~(275) (2020) 1--15.

\bibitem{wang2021sars}
W.~Wang, Z.~Zhou, X.~Xiao, Z.~Tian, X.~Dong, C.~Wang, L.~Li, L.~Ren, X.~Lei,
  Z.~Xiang, et~al., Sars-cov-2 nsp12 attenuates type i interferon production by
  inhibiting irf3 nuclear translocation, Cellular \& Molecular Immunology
  18~(4) (2021) 945--953.

\bibitem{gao2020structure}
Y.~Gao, L.~Yan, Y.~Huang, F.~Liu, Y.~Zhao, L.~Cao, T.~Wang, Q.~Sun, Z.~Ming,
  L.~Zhang, et~al., Structure of the rna-dependent rna polymerase from covid-19
  virus, Science 368~(6492) (2020) 779--782.

\bibitem{tahir2021coronavirus}
M.~Tahir, Coronavirus genomic nsp14-exon, structure, role, mechanism, and
  potential application as a drug target, Journal of Medical Virology 93~(7)
  (2021) 4258--4264.

\bibitem{ogando2020enzymatic}
N.~S. Ogando, J.~C. Zevenhoven-Dobbe, Y.~van~der Meer, P.~J. Bredenbeek, C.~C.
  Posthuma, E.~J. Snijder, The enzymatic activity of the nsp14 exoribonuclease
  is critical for replication of mers-cov and sars-cov-2, Journal of virology
  94~(23) (2020) e01246--20.

\bibitem{ma2015structural}
Y.~Ma, L.~Wu, N.~Shaw, Y.~Gao, J.~Wang, Y.~Sun, Z.~Lou, L.~Yan, R.~Zhang,
  Z.~Rao, Structural basis and functional analysis of the sars coronavirus
  nsp14--nsp10 complex, Proceedings of the National Academy of Sciences
  112~(30) (2015) 9436--9441.

\bibitem{yuen2020sars}
C.-K. Yuen, J.-Y. Lam, W.-M. Wong, L.-F. Mak, X.~Wang, H.~Chu, J.-P. Cai, D.-Y.
  Jin, K.~K.-W. To, J.~F.-W. Chan, et~al., Sars-cov-2 nsp13, nsp14, nsp15 and
  orf6 function as potent interferon antagonists, Emerging microbes \&
  infections 9~(1) (2020) 1418--1428.

\bibitem{yoshimoto2021biochemical}
F.~K. Yoshimoto, A biochemical perspective of the nonstructural proteins (nsps)
  and the spike protein of sars cov-2, The protein journal 40~(3) (2021)
  260--295.

\bibitem{pillon2021cryo}
M.~C. Pillon, M.~N. Frazier, L.~B. Dillard, J.~G. Williams, S.~Kocaman, J.~M.
  Krahn, L.~Perera, C.~K. Hayne, J.~Gordon, Z.~D. Stewart, et~al., Cryo-em
  structures of the sars-cov-2 endoribonuclease nsp15 reveal insight into
  nuclease specificity and dynamics, Nature communications 12~(636) (2021)
  1--12.

\bibitem{vijayan2021structure}
R.~Vijayan, S.~Gourinath, Structure-based inhibitor screening of natural
  products against nsp15 of sars-cov-2 revealed thymopentin and oleuropein as
  potent inhibitors, Journal of proteins and proteomics 12~(2) (2021) 71--80.

\bibitem{deng2018old}
X.~Deng, S.~C. Baker, An `'old'' protein with a new story: Coronavirus
  endoribonuclease is important for evading host antiviral defenses, Virology
  517 (2018) 157--163.

\bibitem{kim2020crystal}
Y.~Kim, R.~Jedrzejczak, N.~I. Maltseva, M.~Wilamowski, M.~Endres, A.~Godzik,
  K.~Michalska, A.~Joachimiak, Crystal structure of nsp15 endoribonuclease
  nendou from sars-cov-2, Protein Science 29~(7) (2020) 1596--1605.

\bibitem{zhang2018structural}
L.~Zhang, L.~Li, L.~Yan, Z.~Ming, Z.~Jia, Z.~Lou, Z.~Rao, Structural and
  biochemical characterization of endoribonuclease nsp15 encoded by middle east
  respiratory syndrome coronavirus, Journal of Virology 92~(22) (2018)
  e00893--18.

\bibitem{guarino2005mutational}
L.~A. Guarino, K.~Bhardwaj, W.~Dong, J.~Sun, A.~Holzenburg, C.~Kao, Mutational
  analysis of the sars virus nsp15 endoribonuclease: identification of residues
  affecting hexamer formation, Journal of molecular biology 353~(5) (2005)
  1106--1117.

\bibitem{mandilara2021role}
G.~Mandilara, M.~A. Koutsi, M.~Agelopoulos, G.~Sourvinos, A.~Beloukas,
  T.~Rampias, The role of coronavirus rna-processing enzymes in innate immune
  evasion, Life 11~(6) (2021) 571.

\bibitem{hackbart2020coronavirus}
M.~Hackbart, X.~Deng, S.~C. Baker, Coronavirus endoribonuclease targets viral
  polyuridine sequences to evade activating host sensors, Proceedings of the
  National Academy of Sciences 117~(14) (2020) 8094--8103.

\bibitem{almazan2006construction}
F.~Almaz{\'a}n, M.~L. DeDiego, C.~Gal{\'a}n, D.~Escors, E.~Alvarez, J.~Ortego,
  I.~Sola, S.~Zuniga, S.~Alonso, J.~L. Moreno, et~al., Construction of a severe
  acute respiratory syndrome coronavirus infectious cdna clone and a replicon
  to study coronavirus rna synthesis, Journal of virology 80~(21) (2006)
  10900--10906.

\bibitem{chang2021nsp16}
L.-J. Chang, T.-H. Chen, Nsp16 2'-o-mtase in coronavirus pathogenesis: possible
  prevention and treatments strategies, Viruses 13~(4) (2021) 538.

\bibitem{sk2020computational}
M.~F. Sk, N.~A. Jonniya, R.~Roy, S.~Poddar, P.~Kar, Computational investigation
  of structural dynamics of sars-cov-2 methyltransferase-stimulatory factor
  heterodimer nsp16/nsp10 bound to the cofactor sam, Frontiers in molecular
  biosciences.

\bibitem{aouadi2017binding}
W.~Aouadi, A.~Blanjoie, J.-J. Vasseur, F.~Debart, B.~Canard, E.~Decroly,
  Binding of the methyl donor s-adenosyl-l-methionine to middle east
  respiratory syndrome coronavirus 2'-o-methyltransferase nsp16 promotes
  recruitment of the allosteric activator nsp10, Journal of virology 91~(5)
  (2017) e02217--16.

\bibitem{wang2015coronavirus}
Y.~Wang, Y.~Sun, A.~Wu, S.~Xu, R.~Pan, C.~Zeng, X.~Jin, X.~Ge, Z.~Shi,
  T.~Ahola, et~al., Coronavirus nsp10/nsp16 methyltransferase can be targeted
  by nsp10-derived peptide in vitro and in vivo to reduce replication and
  pathogenesis, Journal of virology 89~(16) (2015) 8416--8427.

\bibitem{jiang2020repurposing}
Y.~Jiang, L.~Liu, M.~Manning, M.~Bonahoom, A.~Lotvola, Z.-Q. Yang, Repurposing
  therapeutics to identify novel inhibitors targeting 2'-o-ribose
  methyltransferase nsp16 of sars-cov-2.

\bibitem{hazari2022analysis}
R.~Hazari, P.~Pal~Chaudhuri, Analysis of coronavirus envelope protein with
  cellular automata (ca) model, arXiv preprint arXiv:2202.11752.

\bibitem{von1996theory}
J.~Von~Neumann, A.~W. Burks, Theory of self-reproducing automata, University of
  Illinois Press Urbana, 1996.

\bibitem{wolfram1983statistical}
S.~Wolfram, Statistical mechanics of cellular automata, Reviews of modern
  physics 55~(3) (1983) 601--644.

\bibitem{codd1968cellular}
E.~F. Codd, Cellular Automata, Academic Press Inc, New York, 1968.

\bibitem{conway1970game}
J.~Conway, The game of life, Scientific American.

\bibitem{wolfram2002new}
S.~Wolfram, A new kind of science, Wolfram-Media Inc., Champaign, Ilinois,
  United States, 2002.

\bibitem{chaudhuri2018new}
P.~P. Chaudhuri, S.~Ghosh, A.~Dutta, S.~P. Choudhury, A New Kind of
  Computational Biology: Cellular Automata Based Models for Genomics and
  Proteomics, Springer Nature Singapore, ISBN 978-981-1-13-163, 2018.

\bibitem{ppc1}
P.~Pal~Chaudhuri, D.~Roy~Chowdhury, S.~Nandi, S.~Chattopadhyay, Additive
  Cellular Automata -- Theory and Applications, Vol.~1, IEEE Computer Society
  Press, Los Alamitos, USA, ISBN 0-8186-7717-1, 1997.

\bibitem{ridley2015evolution}
M.~Ridley, The Evolution of Everything: How Small Changes Transform Our World,
  Harper Collins UK, ISBN 978-0-00-754247-5, 2015.

\end{thebibliography}

\end{document}